\documentclass[aps,prd,superscriptaddress,reprint,nofootinbib]{revtex4-1}
\usepackage{graphicx}
\usepackage{amsmath}
\usepackage{amssymb}
\usepackage{siunitx}
\usepackage[acronym]{glossaries}
\usepackage{hyperref}
\usepackage{cleveref}
\usepackage{ulem}
\usepackage{xcolor}

\newcommand{\delay}[1]{\mathbf{D}_{#1}}
\newcommand{\adv}[1]{\mathbf{A}_{#1}}

\newacronym{gw}{GW}{gravitational wave}
\newacronym{tdi}{TDI}{Time-Delay Interferometry}
\newacronym{tm}{TM}{test mass}
\newacronym{oms}{OMS}{Optical Metrology System}
\newacronym{lisa}{LISA}{Laser Interferometer Space Antenna}
\newacronym{lpf}{LPF}{LISA Pathfinder}
\newacronym{psd}{PSD}{Power Spectral Density}
\newacronym{asd}{ASD}{Amplitude Spectral Density}
\newacronym{sgwb}{SGWB}{Stochastic Gravitational Wave Background}
\newacronym{ttl}{TTL}{ tilt-to-length }

\begin{document}
\title{On the effectiveness of null TDI channels as instrument noise monitors in LISA}

\author{Martina~Muratore}\email{contact: martina.muratore@aei.mpg.de}
\affiliation{\addressi}
\affiliation{\addressii}
\author{Olaf~Hartwig}
\affiliation{\addres}
\def\addres{SYRTE, Observatoire de Paris-PSL, CNRS, Sorbonne Université, LNE, Paris, France}
\author{Daniele~Vetrugno}
\affiliation{\addressi}
\author{Stefano~Vitale}
\affiliation{\addressi}
\author{William Joseph~Weber}
\affiliation{\addressi}
\def\addressi{Dipartimento di Fisica, Universita di Trento and Trento Institute for 
Fundamental Physics and Application / INFN, 38123 Povo, Trento, Italy}
\def\addressii{Max Planck Institute for Gravitational Physics (Albert Einstein Institute), D-14476 Potsdam, Germany}
\date{\today}

\begin{abstract}
We present a study of the use and limits of the Time-Delay Interferometry null channels for in flight estimation of the Laser Interferometer Space Antenna instrumental noise.  The paper considers how the two main limiting noise sources, test-mass acceleration noise and interferometric phase measurement noise, propagate
through different  Time-Delay Interferometry channels: the Michelson combination X that is the most sensitive to gravitational waves, then the less-sensitive combinations $\alpha$, and finally the null channel $\zeta$. We note that the null channel $\zeta$, which is known to be equivalent to any null channel, not only has a reduced sensitivity to the gravitational waves, but also feature a larger degree of cancellation of the test mass acceleration noise relative to the interferometry noise. This severely limits its use in quantifying the low frequency instrumental noise in the Michelson X combination, which is expected to be dominated by acceleration noise. However, we show that one can still use in-flight noise estimations from $\zeta$ to put an upper bound on the considered noises entering in the X channel, which allows to distinguish them from a strong stochastic gravitational wave background.
 \end{abstract}

\maketitle

\section{Introduction}\label{sec:introduction}

The \gls{lisa} gravitational wave observatory \cite{lisa_proposal} is expected to be continuously dominated by \gls{gw} signals in its \si{\milli\hertz} frequency band. This implies a technical difficulty in quantifying and understanding the instrumental noise of \gls{lisa} in the constant presence of \gls{gw} signals, which is essential for maximizing the observatory scientific return and to identify possible \glspl{sgwb} \cite{lisa_proposal}. 
The \gls{lisa} scientific observables are constructed from so-called \gls{tdi} combinations, which synthesize equal arm interferometers to cancel an otherwise overwhelming contribution from laser frequency noise \cite{Armstrong_1999}. The primary \gls{gw} observables will be obtained from \gls{tdi} channels such as the Michelson X channel. This channel represents a virtual  interferometer with the same principle of measurements of a standard Michelson interferometer, as they are used in ground based \gls{gw} observatories, such as LIGO, Virgo and Kagra\footnote{Technically, LIGO, Virgo and Kagra are Michelson interferometers with Fabry-Perot cavities.}.

In addition to these sensitive channels, it has been shown that so-called null channels can be constructed, which strongly suppress the \gls{gw} signals at low frequencies and might therefore be used for characterizing the instrumental noise \cite{Armstrong_1999, HoganBender2001,muratore2021time}. An example is the null channel $\zeta$, which represents a symmetric measurement across all three arms of the constellation, strongly suppressing its response to \glspl{gw} at low frequencies \cite{Muratore_2020}. It is shown in \cite{muratore2021time} that all possible null channels can be derived from the $\zeta$ channel. Thus, it is sufficient to exclusively focus on $\zeta$ to study the relationship between \gls{tdi} channels sensitive to \glspl{gw} and null channels. The results obtained are then generally applicable to any null channel. 

We notice that, compared to ground-based GW obseratories, a null channel is particularly valuable in \gls{lisa}. While ground-based observatories can exploit correlations between multiple detectors to discriminate the \glspl{gw} from instrumental noise sources, \gls{lisa} will be a single detector, such that the same kind of analysis might not be possible\footnote{In principle, \gls{lisa} is not the only planned space-based \gls{gw} detector targeting \si{\milli\hertz} frequencies, such that there is a possiblity that multiple detectors will be operational at the same time \cite{Hu:2017mde}. However, \gls{lisa} is more advanced in terms of its development compared to other proposed projects, such that this possibility cannot be relied upon. Indeed, \gls{lisa} has been selected in 2017 to be ESA's third large-class mission \cite{lisa_proposal}.
}. Furthermore, as mentioned above, \gls{lisa} is expected to be signal dominated, such that instrumental noise cannot be measured and characterized in flight in isolation from the \gls{gw} signals. \\
Current approaches to the data analysis for \gls{lisa} foresee a "global fit", in which an initial noise model\footnote{A detailed noise model is also essential for the development of the mission, and is already in preparation and anchors the mission hardware requirements.} will be defined to start subtracting resolvable sources in an iterative procedure \cite{Littenberg_2020}. While the background residual noise model is refined in this iterative process, improving the identification of the known sources, the residual noise model does not necessarily distinguish between instrumental and gravitational noise.  As our a priori knowledge of the instrumental noise background is likely to have limited accuracy, this poses a fundamental problem in the identification of a stochastic GW background. 
%
%
%
%
The goal of this paper is to analyze the extent to which the null channel can be used to characterize the dominant noise sources expected to affect the sensitive channels. We then further explore what this implies for the detectability of an isotropic \gls{sgwb} of unknown spectral shape.
 
Following the \gls{lisa} proposal \cite{lisa_proposal}, we consider two general groups of instrumental noise sources: the \gls{tm} acceleration noise and the effective total displacement noise in a one-way single link \gls{tm} to \gls{tm} measurement, which we abbreviate as \gls{oms} noise. As we will discuss in~\cref{sec:instrument-model}, each of these two groups of noise in reality represents a multitude of individual noise contributions driven by different physical effects, both known and possible unknown, such that the exact level and frequency dependence of these \gls{lisa} instrumental noises cannot be reliably calculated a priori. As an example of the difficulty in accurately modeling noise \textit{a priori}, the acceleration noise at \SI{0.1}{\milli\hertz} measured by \gls{lpf}, though compatible with the  \gls{lisa} noise requirements, exceeds by approximately a factor 4 the noise accounted for by the noise model \cite{PhysRevLett.120.061101}. Various key parameters of the  \gls{lisa} noise, including DC residual forces \cite{Armano_2016}, magnetic field gradients \cite{Armano:2020zyx}, residual stray electrostatic fields \cite{article}, optical alignments \cite{PhysRevApplied.14.014030,Hartig_2022}, among others, are all designed to be ideally zero, but with uncertainties that make their residual contribution to the observatory noise both difficult to predict and likely different among the different  \gls{lisa} \gls{tm} or optical readouts.  Other well known noise sources like Brownian noise from gas damping can have a non-trivial time dependence and thus an instantaneous noise  \gls{psd} that is hard to predict \cite{LISAPathfinder:2017yoi}. As such, if noise knowledge is a key factor in extracting  \gls{lisa} science, either for a stochastic background or for noise priors on individual source parameters, then developing \textit{in situ} techniques to quantify the instrument noise is an important task\footnote{It is worth to mention that there are efforts to assess  \gls{tm} acceleration noise by internal measurements. Indeed, some information on \gls{tm} acceleration noise in \gls{lisa} can be obtained by combining position sensing and actuator signals inside a single spacecraft, albeit with a mixing of different degrees of freedom as shown in \cite{https://doi.org/10.48550/arxiv.2202.12735}.}.

\gls{tm} acceleration noise is expected to limit the \gls{gw} sensitive channels, such as the Michelson X, at low frequencies (below a few \si{\milli\hertz}), while the \gls{oms} noise is most relevant for the sensitivity of these channels at high frequencies. On the contrary, we will show that \gls{tm} acceleration noise is effectively suppressed in the $\zeta$ channel, such that it is dominated by \gls{oms} noise at \textit{all} frequencies. This behaviour of the null channels has already been pointed out in \cite{Flauger_2021,https://doi.org/10.48550/arxiv.2201.10902} for the null channel T that is built out of X and the two channels Y and Z (obtained from X by cyclic satellite permutations), considering the case of \gls{lisa} with equal arm-lengths. However, as already shown in \cite{Adams_2010} and  \cite{muratore2021time}, T as a null channel is strongly compromised when the arm lengths are not exactly equal, especially at low frequencies.  
 The null channel $\zeta$ remains less sensitive to \gls{gw} also in the more general unequal arm-lengths scenario. Therefore in the rest of the paper we will discuss only the properties of $\zeta$, and we will simplify the formulas presented to the equal armlength approximation, which has negligable impact on the general conclusions. We will re-introduce the inequality of the arm-lengths when necessary to not bias the computations, and also perform time-domain simulations using realistic orbits to show our equal-armlength models are accurate enough. We discuss in~\cref{sec:discussion,sec:simulation} the impact of our findings, showing that the dominant \gls{oms} noise in the null channel strongly limits its effectiveness for noise characterization in the low frequency regime.
For instance, for a null channel measurement to detect the \gls{tm} acceleration noise relevant to the Michelson X combination at \SI{0.1}{\milli\hertz}, the \gls{oms}  noise would have to be at the  \SI{0.1}{\nano\meter\per\sqrt\hertz} level at \SI{0.1}{\milli\hertz} compared to the requirement of \SI{6}{\nano\meter\per\sqrt\hertz} defined in the proposal \cite{lisa_proposal}, as shown in~\cref{fig:comparison}. Such a low noise level is neither foreseen nor required for \gls{lisa}  \gls{gw}  observation in the TDI Michelson X, Y and Z channels (or the equivalent orthogonal combinations of these, A and E \cite{PhysRevD.66.122002}). Given the inherent uncertainties in the modelling of the noise sources composing the \gls{lisa}  full noise budget pre-flight, we conclude that the null channels can only yield very weak upper limits on the low frequency \gls{tm}  acceleration noise. These constraints on the noise in X become more stringent towards higher frequencies, where both X and $\zeta$ are dominated by \gls{oms}  noise (assuming nominal performance). Note that at very high frequencies, $\zeta$ becomes equally sensitive to X, such that the frequency band in which we can put strong constraints on the instrumental noise in X is rather limited.

The remaining article is divided in four main sections. In~\cref{sec:instrument-model}, we introduce the noise models defined in the \gls{lisa}  proposal and discuss in detail the \gls{tdi}  outputs. In~\cref{sec:discussion} we discuss the use of these null channels to calibrate and measure the instrumental noise during operations and the implication for distinguishing between instrumental noise and \gls{sgwb}.  In~\cref{sec:simulation}, we compare our analytical calculations with time domain simulations using the tools \gls{lisa}  Instrument, \gls{lisa}  GW-Response and PyTDI, which we configured to reflect the noise models given in the \gls{lisa}  proposal~\cite{simulation-model} and our semi-analytical \gls{gw}  response computation. We also study the response of the  significantly less sensitive (compared to X) Sagnac channel $\alpha$, and we explore to which extent the noise entering in  $\alpha$ could be combined with the null channel $\zeta$. 
Finally, we use $\zeta$ to compute an upper limit on the instrumental noise in X, which allows us to identify a strong \gls{sgwb}  in the X data stream.

In the last section we report our conclusion and future perspective.

\section{Instrumental noise modeling and TDI outputs\label{sec:instrument-model}}

In this section, we briefly introduce the main limiting noise sources left after  TDI suppression of laser frequency noise, and any possible further subtraction of any known calibrated and measured instrumental noise sources, such as, for example, the optical tilt to length cross-coupling to spacecraft motion (\cite{Hartig_2022} and \cite{PhysRevApplied.14.014030}). 

The remaining noises, for which we have neither a measurement for coherent subtraction nor a high precision a priori model, falls into two broad categories \cite{lisa_proposal}, the acceleration noise of each individual \gls{tm} and an overall optical metrology noise term for each single link measurement. For the \gls{gw}  sensitive \gls{tdi}  channels, the former is expected to be the limiting noise source at low frequencies, while the latter is most relevant at high frequencies. We then compute how these noise sources propagate through different \gls{tdi}  channels, and discuss to which extent and at which frequencies the null channel can be a useful noise monitor for the \gls{gw}  sensitive channels. 

We will express the phase outputs measured by \gls{lisa}, used to build the \gls{tdi} channels, as an effective displacement signal in units of meters. 
\subsection{TM acceleration noise}
In the assumption of a perfect spacecraft jitter subtraction, we can ignore the complications of the split interferometry scheme \cite{lisa_proposal} and assume for the calculation of the \gls{tdi}  outputs that the two optical-benches, say $\mathrm{OB}_i$ and $\mathrm{OB}_j$, were two free-falling particles that accelerate along their relative line of sight towards each other with accelerations $g_i$ and $g_j$ respectively. $g_i$ and $g_j$ describe the \glspl{tm}  acceleration noise with respect to the local inertial frame, that for \gls{lisa}, can be associated with the one defined by the incoming laser beam (See~\cref{ssec:tm-noise}).

We will denote the overall \gls{tm}  acceleration noise \gls{psd}  of a single \gls{tm}  as $S_{g_{ij}}$. We assume for simplicity that all \gls{tm}  acceleration noises are fully uncorrelated to each other, although this might not be the case in reality\footnote{For example, both \gls{tm}  inside one spacecraft might be affected by common-mode effects such as temperature fluctuations or tilt-to-length couplings due to rotation of the spacecraft.}. Furthermore, in our assumptions of free-falling optical benches, we can directly convert the acceleration noise of a single \gls{tm}  to an equivalent displacement of the correspondent optical bench, whose PSD is given as
\begin{equation}
	S_{g_{ij}}^{disp} =S_{g_{ij}} /(2 \pi f)^4.
\end{equation}
We will denote the time series associated with this displacement as $x^g_{ij}(t)$.

Note that the exact noise shape and amplitude of each individual $S_{g_{ij}}$ will result from the superposition of a multitude of physical effects. While many of these effects have been characterized during the very successful \gls{lpf}  mission, the total measured noise is considerably larger than the sum of these known sources, indicating the difficulty in achieving a complete, accurate model~\cite{PhysRevLett.120.061101,lpf_glitch2022}. We therefore cannot assume to have accurate prior knowledge of the overall acceleration noise, and would need to rely on in-flight measurements to constrain its value for each \gls{tm}.

\subsection{Optical Metrology System noise}

We summarize as \gls{oms} noise any imperfection in the ability of the \gls{oms} to determine the separation between two \glspl{tm}  in a single link. \\ Similar to the overall acceleration noise acting on each \gls{tm}, the overall \gls{oms}  noise affecting a single link will be a superposition of many physical effects. In addition, the overall \gls{oms} noise summarizes noise entering due to different instrumental subsystems, such as the telescope, optical bench, phase measurement system, laser, clock and \gls{tdi} processing  \cite{lisa_proposal}. Note that some of these noise sources can again be correlated (similar to the \gls{tm} acceleration noise), while here we assume they are not.

For simplicity, we consider only a single uncorrelated noise term in each single link measurement, whose  \gls{psd} we denote by $S_{oms_{ij}}(f)$. We will denote the time series of these single link \gls{oms} as $x^m_{ij}(t)$.

We remark that while the  \gls{tm} acceleration was measured by \gls{lpf} in realistic flight conditions, we expect new challenges and additional uncertainty in the pre-flight characterization of the \gls{oms}. While many terms in the \gls{oms} budget are well calculated from models and ground testing (such as shot noise and phasemeter noise), the end-to-end inter-spacecraft \gls{lisa} optical measurement has never been performed and we can expect unknowns. This is especially true for the low frequency regime, as there will likely be no possible on-ground long term testing before flight ($>$ 1 month). We also cannot assume this noise to be stationary over the mission duration, as we must expect that some of the physical parameters governing its level will change during the mission duration, such as \gls{ttl} effects \cite{Hartig_2022}.

\subsection{Analytical calculation of noise couplings into TDI X, \texorpdfstring{$\alpha$}{alpha} and \texorpdfstring{$\zeta$}{zeta}}
In this section we illustrate the transfer of \gls{tm} acceleration and \gls{oms} noises into the relevant \gls{tdi} variables with simplified expressions valid in the low frequency limit (for angular frequencies $\omega \ll 1/\tau $).  The full time dependence will be used in the calculations that follow in~\cref{sec:simulation}.

We derive these noise couplings in the assumption that the light propagation time across all three \gls{lisa} arms is equal to the same value $\tau\approx\SI{8.3}{\second}$. As discussed in~\cref{ssec:tm-noise}, the two noise sources we consider enter into a single link as
\begin{equation}
	\eta_{ij}(t) = x^g_{ji}(t - \tau) + x^g_{ij}(t) + x^m_{ij}(t).\label{eq:link}
\end{equation}
Here, $\eta_{ij}(t)$ represents the so-called intermediary \gls{tdi} variables representing the single link \gls{tm} to \gls{tm} measurement. The first index $i$ represents the spacecraft the measurement is performed on at time $t$, while the second index $j$ denotes the distant spacecraft light was emitted from at time $t - \tau$.

From these measurements it is possible to build the \gls{tdi} channels  X, $\alpha$ and $\zeta$ as shown in~\cref{tablecombinations1}.   The table makes use of the of time shift operators which act on time dependent functions by evaluating them at another time, see~\cref{sec:notations}. \\

\begin{table*}
\centering
\resizebox{\textwidth}{!}{
\renewcommand{\arraystretch}{1.3}
\begin{tabular}{| c | c |}
\hline
Name & Expression \\
\hline
$\alpha$ & $ \left(1-\delay{13}\delay{32}\delay{21}\right)\eta_{12}+\left(\delay{12}-\delay{13}\delay{32}\delay{21}\delay{12}\right)\eta_{23}+\left(\delay{12}\delay{23}-\delay{13}\delay{32}\delay{21}\delay{12}\delay{23}\right)\eta_{31}$ \\
& $-\left(1-\delay{12}\delay{23}\delay{31}\right)\eta_{13}-\left(\delay{13}-\delay{12}\delay{23}\delay{31}\delay{13}\right)\eta_{32} -\left(\delay{13}\delay{32}-\delay{12}\delay{23}\delay{31}\delay{13}\delay{32}\right)\eta_{21}$
\\ \hline 
$\zeta$ & $\left(\delay{32}\delay{23}\adv{31}-\delay{31}\adv{12}\delay{23}\adv{31}\right)(\eta_{13} - \eta_{12}) + \left(1-\delay{32}\delay{23}\adv{31}\delay{12}\adv{23}\right)(\eta_{31}-\eta_{32})$ \\ 
& $ + \left(\delay{31}\adv{12}\delay{23}\adv{31}\delay{12} - \delay{31}\adv{12}\right)\eta_{21} - \left(\delay{32}-\delay{31}\adv{12}\right)\eta_{23}$\\ 
\hline
$X$ & $ \left(1-\delay{13}\delay{31}-\delay{13}\delay{31}\delay{12}\delay{21}+\delay{12}\delay{21}\delay{13}\delay{31}\delay{13}\delay{31}\right)\eta_{12}- \left(1-\delay{12}\delay{21}-\delay{12}\delay{21}\delay{13}\delay{31}+\delay{13}\delay{31}\delay{12}\delay{21}\delay{12}\delay{21}\right)\eta_{13} $ \\ 
& $+ \left(\delay{12}-\delay{13}\delay{31}\delay{12}-\delay{13}\delay{31}\delay{12}\delay{21}\delay{12}+\delay{12}\delay{21}\delay{13}\delay{31}\delay{13}\delay{31}\delay{12}\right)\eta_{21}- \left(\delay{13}-\delay{12}\delay{21}\delay{13}\delay{31}\delay{13}+\delay{13}\delay{31}\delay{12}\delay{21}\delay{12}\delay{21}\delay{13}-\delay{12}\delay{21}\delay{13}\right)\eta_{31}$\\ \hline 
\end{tabular}}
\renewcommand{\arraystretch}{1}
\caption{\label{tablecombinations1} List of the TDI $\alpha$, $\zeta$ and X as given in \cite{muratore2021time}, expressed in terms of time shifts applied to the intermediary TDI variables $\eta_{ij}$. The table has been adapted from \cite{hartwig2022characterization}.}
\end{table*}
As a preliminary analysis of the usefulness of $\zeta$ for noise characterization, it is instructive to consider the expression for $\zeta$ in~\cref{tablecombinations1} in the equal-armlength limit, where it simplifies to
\begin{equation}
\zeta = (1 - D)\left(\eta_{12} - \eta_{13} + \eta_{23} - \eta_{21} + \eta_{31} - \eta_{32}\right),\label{eq:zeta_TMcorr}
\end{equation}
with $D$ as the equal-arm delay operator.

We observe that $\zeta$ is insensitive to noise which is correlated such that it enters both of the two single-link measurements recorded on-board a single spacecraft in exactly the same way\footnote{Inspecting~\cref{tablecombinations1}, this cancellation is exact for spacecraft 1 and 3 regardless of any assumptions on the delays, while equal noise terms in $\eta_{23}$ and $\eta_{21}$ only cancel in the assumption of a constellation with 3 constant (but possibly unequal) arms. Note that such noise on spacecraft 2 will still be strongly suppressed considering realistic orbits.}. More generally, noise entering correlated (but not exactly equal) in the two measurements, such as noise in the measurements $\eta_{12}$ and $\eta_{13}$  in Eq. \ref{eq:zeta_TMcorr}, will be suppressed, while measurement noise entering anti-correlated will be amplified with respect to the uncorrelated case.

Considering the expressions for $\alpha$ and X in~\cref{tablecombinations1} in the equal-armlength limit, we see that this is not the case for these channels, where the spacecraft links enter asymmetrically, and equal noise terms do not cancel in the same way. This means $\zeta$ cannot be used to characterize noise with these correlation properties. Furthermore, as we will discuss in the following,  noise entering correlated in the two directions of a link (such as the \gls{tm} acceleration noises) will also be suppressed in $\zeta$ with respect to noise which is fully uncorrelated in each link.

\subsubsection{Analytical computation of the acceleration noise for the TDI X, \texorpdfstring{$\alpha$}{alpha}, \texorpdfstring{$\zeta$}{zeta}}
Assuming equal arm lengths, we find that the \gls{tm} acceleration noise for the combinations X, $\alpha$ and $\zeta$ can be approximated as:
\begin{subequations}\label{eq:acc-tdi-approx}
\begin{align}\label{eq:X_apprx}
X_{g}( t )  & \approx  16 \tau ^2 \left({x^g_{12}}''(t)-{x^g_{13}}''(t)+{x^g_{21}}''(t)-{x^g_{31}}''(t)\right), \\ 
\begin{split} \label{eq:alpha_apprx}
\alpha_{g}( t ) & \approx 3 \tau ^2 \Big(3 {x^g_{12}}''(t)-3 {x^g_{13}}''(t)+{x^g_{21}}''(t)+{x^g_{23}}''(t) \\
& \qquad \qquad -{x^g_{31}}''(t)-{x^g_{32}}''(t)\Big), 
\end{split} 
\\
\begin{split}  \label{eq:zeta_apprx}
\zeta_{g}( t ) &  \approx  \tau ^2 \Big({x^g_{12}}''(t)-{x^g_{13}}''(t)-{x^g_{21}}''(t)+{x^g_{23}}''(t) \\
& \qquad \qquad +{x^g_{31}}''(t)-{x^g_{32}}''(t)\Big).
\end{split} 
\end{align}
\end{subequations}
where we have expanded to leading order in the average light travel time $\tau$. This expansion is only valid at timescales much greater than $\tau\approx \SI{8.3}{\second}$.

They allow us to see immediately which \glspl{tm} dominate the noise for each of the \gls{tdi} combinations.  The full expressions without expansion can be found in~\cref{a:acc}.

Under these assumptions, we can see that for the \gls{tm} acceleration noise, the \gls{tdi} combination $\zeta$ measures a signal that is a combination of all the 6 \glspl{tm}. Similarly, $\alpha$ also measures a combination of all the six \glspl{tm} but with different coefficients.  The Michelson X measures instead a combination of only four \glspl{tm}.

\subsubsection{\label{sec:disp}Analytical computation of the metrology noise for the TDI X, \texorpdfstring{$\alpha$}{alpha} and \texorpdfstring{$\zeta$}{zeta}}
Following the same steps as in the previous section we find that the propagation of the \gls{oms} noise through the different \gls{tdi} variables is:
\begin{subequations}\label{eq:oms-tdi-approx}
\begin{align}\label{Eq:Xoms_approx}
X_{oms}(t) & \approx  8 \tau ^2 \left({x^{m}_{12}}''(t)-{x^{m}_{13}}''(t)+{x^{m}_{21}}''(t)-{x^{m}_{31}}''(t)\right), \\ 
\begin{split} \label{Eq:alphaoms_approx}
\alpha_{oms}(t) & \approx 3 \tau  \Big({x^{m}_{12}}'(t)-{x^{m}_{13}}'(t)-{x^{m}_{21}}'(t)+{x^{m}_{23}}'(t) 
\\ 
& \qquad \qquad + {x^{m}_{31}}'(t)-{x^{m}_{32}}'(t)\Big), 	
\end{split}
\\
\begin{split}  \label{Eq:zetaoms_approx}
\zeta_{oms} (t)  & \approx  \tau  \Big({x^{m}_{12}}'(t)-{x^{m}_{13}}'(t)-{x^{m}_{21}}'(t)+{x^{m}_{23}}'(t)
\\
& \qquad \qquad +{x^{m}_{31}}'(t)-{x^{m}_{32}}'(t)\Big).
\end{split}
\end{align}
\end{subequations}
Here, we expanded again to leading order in the average light travel time $\tau$ to see what the contributions of the \gls{oms} noise for each of the \gls{tdi} combinations are at low frequency. This expansion is therefore only valid at timescales much greater than $\tau\approx \SI{8.3}{\second}$.
The full expressions for the \gls{oms} noise of the aforementioned combinations without expansion can again be found in~\cref{a:acc}.

We see that for the \gls{oms} noise, at low frequency, $ \alpha_{oms}( \tau)\approx 3 \zeta_{oms}( \tau)$ \footnote{The results that $\zeta \approx 3 \alpha$ at low frequencies for \gls{oms} noise could also be seen from the relationship between $\zeta$ and $\alpha$, $\beta$, $\gamma$ known from the literature. I.e., it is known for the first generation variables that (in the equal armlength limit) $(1 - D^3)\zeta = (D-D^2)(\alpha + \beta + \gamma)$, where $D$ denotes a time-shift by $\tau$. In the low-frequency expansion, this becomes $3 \zeta' \simeq \alpha'+\beta'+\gamma' \simeq 3 \alpha'$, where the last approximation is only valid for the \gls{oms} noise terms, which enter identically in the first generation $\alpha$, $\beta$, $\gamma$ and $\zeta$ (up to delays). For the second generation variables considered here, as shown in \cite{hartwig2022characterization}, $\zeta$ receives an extra factor $(1 - D)$, while $\alpha$, $\beta$, $\gamma$ instead receive a factor $(1 - D^3)$, such that overall, we have $\zeta''  \simeq 3 \alpha''$.\label{foot:3alphazeta}}. Furthermore, we observe that the \gls{oms} noise enters $\alpha$ and $\zeta$ only as a first derivative, while the \gls{tm} acceleration noise entered as a second derivative. This reflects a low frequency suppression of \gls{tm} acceleration noise terms relative to  \gls{oms} noise terms, due to the difference between the two single link measurements (see~\cref{eq:link,tablecombinations1}) along the same arm, which contain the same  \gls{tm} acceleration noise terms but with different delays. For the TDI X, on the other hand, both  \gls{oms} noise and  \gls{tm} acceleration noise enter as a second derivative (compare \cref{eq:X_apprx} vs. \cref{Eq:Xoms_approx}).

\section{Preliminary discussion\label{sec:discussion}}
\begin{figure}
    \centering
    \includegraphics[width=\columnwidth]{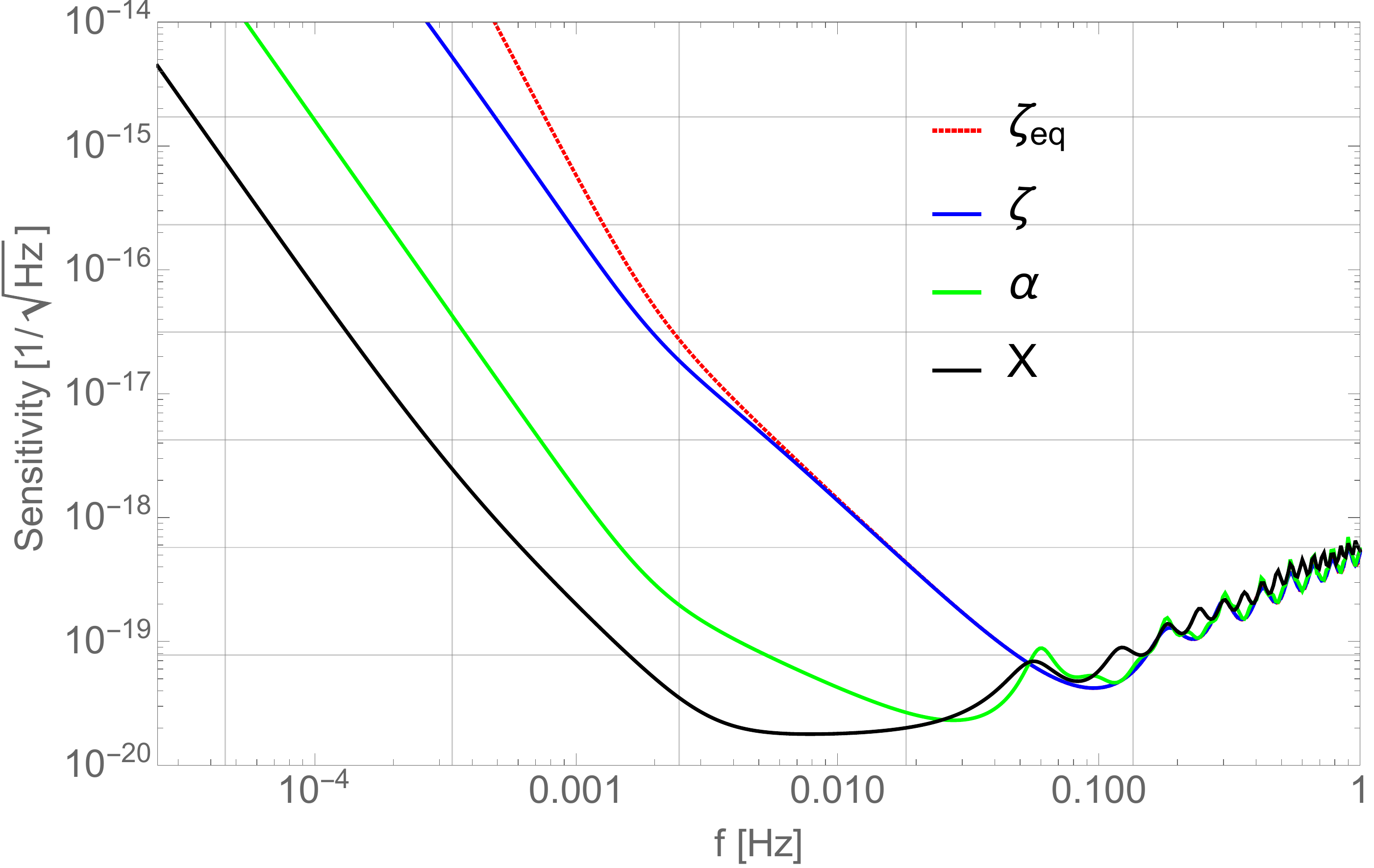}
    \caption{Gravitational wave strain noise spectral density calculation for TDI combinations X, $\alpha$, $\zeta$ averaged over sky position and polarization (see~\cref{sensitivitycom}). The sensitivity are computed considering equal armlength for X and $\alpha$, while for $\zeta$ we also include the sensitivity for three unequal fixed armlength. \label{fig:sens}}
\end{figure}

As known from the literature \cite{Armstrong_1999}, and also shown in~\cref{fig:sens}, the Michelson X channel is sensitive to  \glspl{gw}. One of the expected  \gls{gw} sources for  \gls{lisa} is the  \gls{sgwb}, which in principle could be observed across the whole frequency band \cite{LISASC}. Such a  \gls{sgwb} will be superimposed with the instrumental noise entering in the X channel, such that we should measure an excess in noise power  with respect to the real instrumental noise in order to detect a \gls{sgwb}. However, as discussed in~\cref{sec:instrument-model}, we cannot rely on noise modelling and on-ground testing to fully characterize the instrumental noise, such that we would need to measure it in-flight.  \\

One option for measuring the instrumental noise would be to consider the output of a null channel like $\zeta$ which, at least at low frequencies, is insensitive to \glspl{gw}  \cite{HoganBender2001,Armstrong_1999,muratore2021time}. \Cref{fig:sens} shows the sensitivities for the TDI X, $\alpha$ and $\zeta$, computed as described in~\cref{sensitivitycom}. For X and $\alpha$, we find the sensitivity to be unaffected by an armlength mismatch, while $\zeta$ becomes slightly more sensitive to \glspl{gw} when considering three unequal constant arms\footnote{We followed Ref. \cite{muratore2021time} to estimate the light travel time in case of three unequal constant arms.} instead of three equal constant arms. Nevertheless, in both cases $\zeta$ remains less sensitive than X by many orders of magnitude, such that we will consider the simpler equal armlength case for computing the noises and \gls{gw} response of the \gls{tdi} variables in the following. \\
We can do a preliminary calculation by computing the total noise  \glspl{psd} for  \gls{tdi} X and $\zeta$, which we denote as $S^{noise}_X$ and $S^{noise}_\zeta$. We compute them as the linear sum of the \gls{oms} and  \gls{tm} acceleration noises, respectively, using the low-frequency expansions given in~\cref{eq:acc-tdi-approx,eq:oms-tdi-approx}. 
We get
\begin{align}
S^{noise}_X & \approx 64 \tau^4 \omega^4 \bigg{(}4\sum_{ij \in \mathcal{I}_X} S_{g_{ij}}^{disp} +  \sum_{ij\in \mathcal{I}_X} S_{oms_{ij}}\bigg{)}\label{eq:S_X},\\
S^{noise}_{\zeta} & \approx \tau^2 \omega^2 \bigg{(}\tau^2\omega^2 \sum_{ij\in\mathcal{I}_\zeta}  S_{g_{ij}}^{disp}+  \sum_{ij\in\mathcal{I}_\zeta} S_{oms_{ij}}\bigg{)}\label{eq:S_zeta},
\end{align}
as the overall noise entering in the two channels, valid for $\omega \tau \ll 1$. Here, we introduced the index sets $\mathcal{I}_X = \{ 12,21,13,31 \}$ and $\mathcal{I}_\zeta = \{ 12,23,31,13,32,21 \}$ for the four and six optical links (received at spacecraft i from spacecraft j) and  \gls{tm} acceleration noise terms ( \gls{tm} in spacecraft i accelerated towards spacecraft j) appearing in X and $\zeta$, respectively. 

We can observe that a  \gls{tm} displacement due to  \gls{tm} acceleration noise and the \gls{oms} noise enter with almost the same transfer function into the X channel, up to an additional factor 4 in the \gls{tm} displacement. 
Conversely, in $\zeta$ the \gls{tm} acceleration noise is suppressed towards low frequencies by a factor $\tau^2 \omega^2$ relative to the \gls{oms} noise. This implies that while  \gls{tm} acceleration noise becomes dominant in X for frequencies in which (on average) $S_{oms} \ll 4 S^{disp}_g $, for $\zeta$ the same holds only if $S_{oms}\ll \tau^2\omega^2 S^{disp}_g$. Considering frequencies in the range \SIrange{e-3}{e-4}{\hertz}, the  \gls{tm} acceleration noise pre-factor $\tau^2\omega^2 $ (cf. \cref{eq:S_zeta})  is between \numrange{2.7e-5}{2.7e-3}. This means that the  \gls{oms} noise would have to be from ten parts in a million to one part in a thousand smaller in power than the  \gls{tm} acceleration noise in order for the latter to have the same order of magnitude as the OMS noise in the $\zeta$ channel in the sub-\si{\milli\hertz} band.\\
 As such, the null channels ability to monitor noise in the  \gls{gw} sensitive channels at low frequencies is limited. $\zeta$ could only be used to reliably detect the relevant sub-mHz noise in a worst case scenario where the  \gls{tm} acceleration noise is orders of magnitude larger than the  \gls{oms} noise in these frequency ranges, such that it overcomes the scaling factor $\tau^2\omega^2 $ and becomes dominant in both $\zeta$ and X. 
 
 As we will discuss in the next section, the currently assumed requirements for  \gls{tm} acceleration and  \gls{oms} noises are very far away from these values. Nevertheless, we can still formulate upper and lower bounds on a  \gls{sgwb} signal based on X and $\zeta$ for the full  \gls{lisa} frequency band. 
\section{Upper limits, expected noise levels and simulations\label{sec:simulation}}
\begin{figure*}
	\centering
	\includegraphics[width=\textwidth]{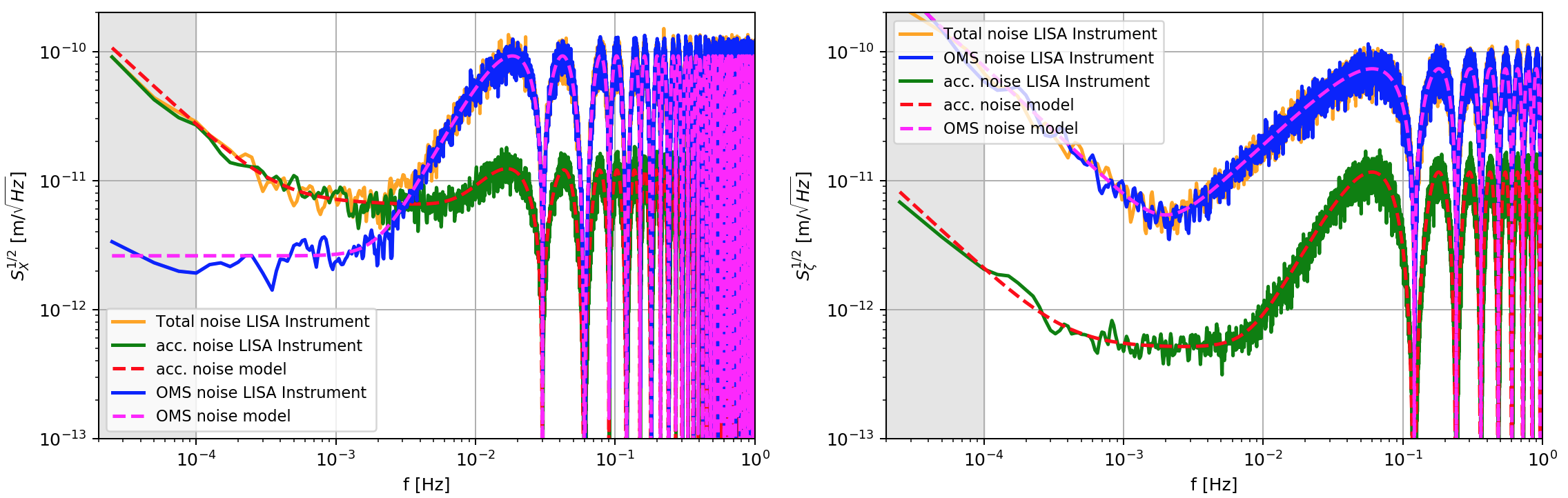}
	\includegraphics[width=0.7\textwidth]{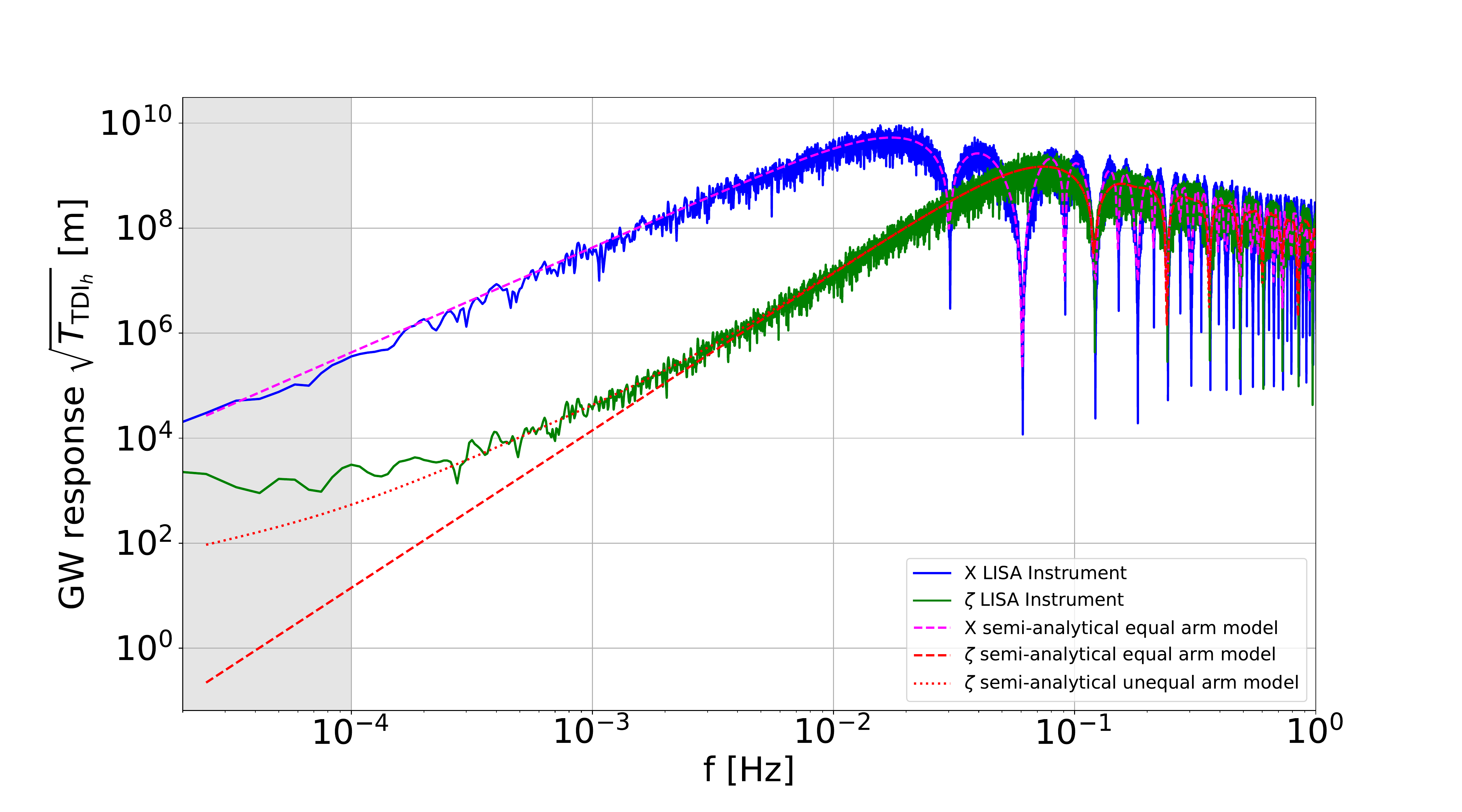}
	\caption{Upper plots: the left one shows the PSD of the TDI X and the right one the PSD of $\zeta$ for the TM acceleration noise, the optical metrology noise and the total noise as simulated with LISA Instrument compared with the respective analytical models. Lower plot: response to GW for TDI X and $\zeta$ as simulated with LISA Instrument and LISA GW-Response compared with the semi-analytical models computed considering equal armlength for X and both equal and three fixed unequal armlength for $\zeta$ \label{fig:psdstotal} as described in~\cref{sensitivitycom}.}
\end{figure*}
After the preliminary analysis in~\cref{sec:discussion}, let us now drop the low-frequency approximation and discuss the accuracy to which we can use X, $\alpha$ and $\zeta$ to identify a potential  \gls{sgwb} with  \gls{lisa}. 

To this end, we briefly introduce the currently assumed noise levels given in the literature \cite{BPH2021}. Note that these should be thought of as the performance requirements we aim to reach with as much margin as possible, not as accurate predictions of the actual in-flight performance. \\ We also perform time domains simulations using  \gls{lisa} Instrument \cite{lisa-instrument} and pyTDI \cite{pytdi} to test our expressions for how these noises couple into the different  \gls{tdi} variables. Similarly, we also perform time domain simulations to test our semi-analytical computation of the  \gls{gw} response of different  \gls{tdi} variables presented in~\cref{sensitivitycom}, using the tool GW-response~\cite{gw-response}. Using simulations allows us to compare our (semi)-analytical expressions, computed assuming equal arm-lengths, with data generated using realistic  \gls{lisa} orbits provided by ESA. 
\subsection{Analytical model and simulations\label{ssec:analytical-model-and-simulations}}
\subsubsection{Instrumental Noise}
Considering the analytical computation in time domain of the  \gls{tm} acceleration and  \gls{oms} noises for the  \gls{tdi} X, $\alpha$, $\zeta$ in~\cref{a:acc}, we can estimate the  \gls{psd} of the aforementioned  \gls{tdi} combinations assuming all  \gls{tm} acceleration and  \gls{oms} noises to be uncorrelated, which yields
\begin{widetext}
	\begin{subequations}
		\begin{align}
			\label{eq:S_gX}
			S_{X_g}  = &\underbrace{256 \sin ^4(\tau  \omega ) \cos ^2(\tau  \omega )}_{T_{X_g}} \left(\left(S_{g_{12}}^{disp}+S_{g_{13}}^{disp}\right) \cos ^2(\tau  \omega
			)+S_{g_{21}}^{disp}+S_{g_{31}}^{disp}\right),\\
			\label{eq:S_galpha}
			S_{\alpha_g}  = &  16\sin^2\left(\frac{\tau  \omega }{2}\right)\sin^2\left(\frac{3\tau  \omega }{2}\right) \left((1 + 2\cos(\tau\omega))^2\left(S_{g_{12}}^{disp} +S_{g_{13}}^{disp}\right)  +S_{g_{21}}^{disp}+S_{g_{23}}^{disp}+S_{g_{31}}^{disp}+S_{g_{32}}^{disp}\right),\\
			\label{eq:S_gzeta}
			S_{\zeta_g} = & \underbrace{16 \sin ^4\left(\frac{\tau  \omega	}{2}\right)}_{T_{\zeta_g}} \left(S_{g_{12}}^{disp}+S_{g_{13}}^{disp}+S_{g_{21}}^{disp}+S_{g_{23}}^{disp}+S_{g_{31}}^{disp}+S_{g_{32}}^{disp}\right),
		\end{align}
	\end{subequations}
and
\begin{subequations}
\begin{align}
	\label{eq:S_dispX}
	S_{X_{oms}}  =& \underbrace{64 \sin ^4(\tau \omega ) \cos ^2(\tau \omega )}_{T_{X_{oms}}}\left(S_{\text{oms}_{12}}+S_{\text{oms}_{13}}+S_{\text{oms}_{21}}+S_{\text{oms}_{31}}\right) ,\\
		\label{eq:S_dispalpha}
	S_{\alpha_{oms}}  =&4 \sin	^2\left(\frac{3 \tau \omega }{2}\right) \left(S_{\text{oms}_{12}}+S_{\text{oms}_{13}}+S_{\text{oms}_{21}}+S_{\text{oms}_{23}}+S_{\text{oms}_{31}}+S_{\text{oms}_{32}}\right) ,\\
	\label{eq:S_dispzeta}
	S_{\zeta_{oms}} =& \underbrace{ 4 \sin^2\left(\frac{\tau \omega}{2}\right)}_{T_{\zeta_{oms}}} \left(S_{\text{oms}_{12}}+S_{\text{oms}_{13}}+S_{\text{oms}_{21}}+S_{\text{oms}_{23}}+S_{\text{oms}_{31}}+S_{\text{oms}_{32}}\right).
\end{align}
\end{subequations}
We verify the validity of these equations (derived in the equal-arm limit) using time domain simulations with realistic orbits. We disabled all noise sources available in  \gls{lisa} Instrument except  \gls{tm} acceleration noise and  \gls{oms} noise in the inter-spacecraft interferometer, and set all noises of the same type to the same level, as given in \cite{BPH2021}. 

For the  \gls{tm} acceleration noises, this means a value of
\begin{equation}
	S_{g_{ij}}(f) = \left(3 \times 10^{-15}~\frac{\textrm{m}}{\textrm{s}^2~\sqrt{\textrm{Hz}}}\right)^2 \times \left(1 + \left(\frac{0.4 ~ \textrm{mHz}}{f}\right)^2\right)\left(1 + \left(\frac{f}{8 ~ \textrm{mHz}}\right)^4\right),
\end{equation}
\end{widetext}
which translates to
\begin{equation}\label{eq:tmacc}
	S_{g_{ij}}^{disp}(f) = S_{g_{ij}}(f) /(2 \pi f)^4
\end{equation}
in terms of displacement.

The noise level of the   \gls{oms} is instead given as 
\begin{equation}
	S_{oms_{ij}}(f) =\left(15 ~\textrm{pm}/\sqrt{\textrm{Hz}}\right)^2 \times \left(1 + \left(\frac{2~\textrm{mHz}}{f}\right)^4\right)\label{eq:readout},
\end{equation}
where the factor $1 + \left(\frac{2~\textrm{mHz}}{f}\right)^4$ is a low frequency relaxation term introduced to take into account our difficulties in measuring that noise below a few mHz from on-ground laboratory experiments. This relaxation is further justified by the fact that it has no impact on the low frequency  \gls{gw} sensitivity in X, as  \gls{oms} noise remains very subdominant in X compared to  \gls{tm} acceleration noise even when including this factor, as visible in the left plot in~\cref{fig:psdstotal}. Note that the estimated  \gls{oms} noise model for  \gls{lpf} also includes a low-frequency relaxation to account for possible thermally driven effects~\cite{PhysRevLett.126.131103}. However, these were likely buried in the \gls{lpf} noise at lower frequencies where \gls{tm} acceleration noise is believed to dominate. The upper part of~\cref{fig:psdstotal} shows the results of three simulation runs with \gls{lisa} Instrument where we enable either one, the other or both of these noise sources. We use PyTDI to compute the Michelson X and $\zeta$ variables.

First, we note that in all cases, the simulated data, with realistic, unequal arm orbits, agrees well with the simplified equal-arm analytic expressions derived for the noise.  We see that in the $\zeta$ channels the \gls{oms} noise is dominant over the \gls{tm} acceleration noise at all frequency, while \gls{tm} acceleration noise becomes the dominant noise source for X below a few \si{\milli\hertz}.

\begin{figure}
    \centering
        \includegraphics[width=\columnwidth]{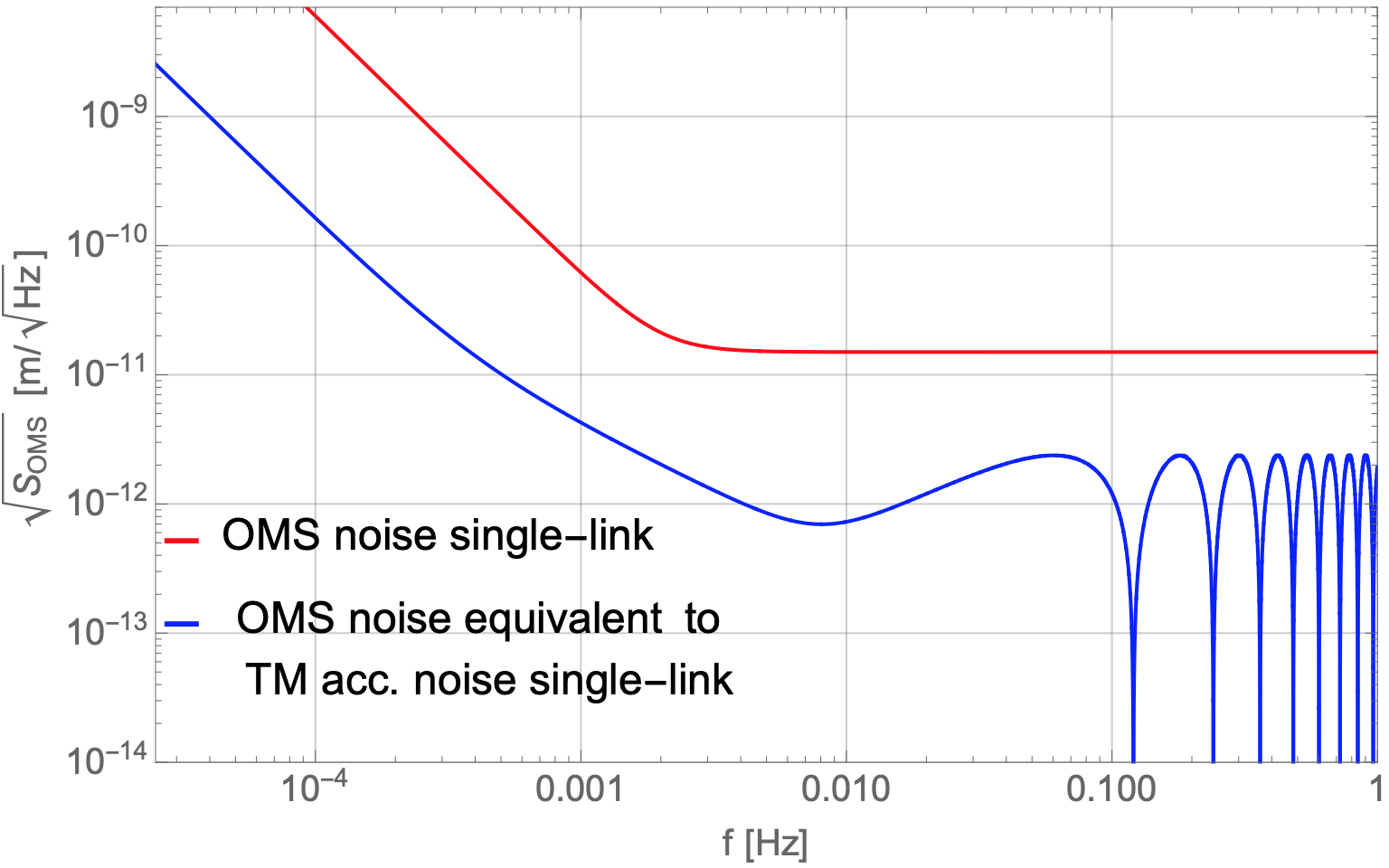}
    \caption{Comparison between the ASD of the optical metrology noise given in \cite{BPH2021} and the equivalent metrology noise in a single LISA link which would be required for the test mass acceleration noise to be dominant in the null channel $\zeta$.\label{fig:comparison}}
\end{figure}

Moreover, if we assume all noises of the same type to have the same noise level, we can use~\cref{eq:S_dispzeta,eq:S_gzeta} to compute that for $\zeta$ we would need an \gls{oms} noise level of $S_{\zeta_{oms}} = 24 \sin^2(\frac{\tau \omega}{2})  \times 4 \sin^2(\frac{\tau \omega}{2}) S_{g}^{disp}$  such that \gls{oms} and \gls{tm} acceleration noises appear at the same magnitude. This can be translated in the single link \gls{oms} noise contribution with the value of $4 \sin^2(\frac{\tau \omega}{2}) S_{g}^{disp} $, which we compare in~\cref{fig:comparison} to the requirement for the \gls{oms} noise given in~\cref{eq:readout}.

We observe that this noise level is likely impossible to achieve as the new required level of \gls{oms} noise is \SI{160}{\pico\meter\per\sqrt\hertz} at \SI{0.1}{\milli\hertz}, orders of magnitude below the currently assumed value. It must be also kept in mind that this conclusion is true keeping fixed the \gls{tm} acceleration noise level to the nominal value, while drastically lowering the \gls{oms} noise level. Any improvement of the \gls{tm} acceleration noise in \gls{lisa} would make the upper limit achieved by the null channel even less relevant.
However, \cref{fig:psdstotal} shows that, at least assuming nominal noise levels, both X and $\zeta$ are dominated by \gls{oms} noise above 4$~\textrm{mHz}$, which might suggest that $\zeta$ can put a stronger constraint on the instrumental noise in this frequency range. \\
\subsubsection{Gravitational wave response}

We denote the \gls{psd} of the X and $\zeta$ channels due to \glspl{gw} as
\begin{equation}\label{eq:def-Sxh}
	S_{X_h} = T_{X_h} S_h, \qquad  S_{\zeta_h} = T_{\zeta_h} S_h.
\end{equation}
 $S_h$ is expressed as a dimensionless stochastic \gls{gw} strain in \si{\per\hertz}, while $S_{X_h}$ and $S_{\zeta_h}$ are expressed in \si{m\squared\per\hertz}. Therefore, the response functions $T_{X_h}$ and $T_{\zeta_h}$ each include a conversion factor that has units \si{\meter\squared}.

The lower plot in~\cref{fig:psdstotal} shows the \gls{gw} responses to a  \gls{sgwb}  of  \gls{tdi} X and $\zeta$ for 51 stochastic \gls{gw} sources isotropically distributed over the sky, computed in the frequency domain as described in~\cref{sensitivitycom}. To verify the validity of these equal arm-length models, we compare them to time domain simulations using the tools \gls{lisa} Instrument, GW-response and PyTDI which use realistic ESA orbits. We inject an isotropic  \gls{sgwb} computed from $N = 48$ sources into the time-domain simulation\footnote{The GW-response tool used to compute the time domain response only allows certain fixed numbers of stochastic sources. $N=48$ is the closest valid value to what we used in the Fourier domain computation.} and disable all instrumental noises. The two strain time series $h_+$, $h_\times$ for each source are computed as a white noise of amplitude $1/N$, where N is the number of  \gls{gw} sources, to overall simulate a sky-averaged response to a unit amplitude \gls{sgwb}. \\
	 
We see that for X our model for equal arms agrees with the simulations, while the equal arm-length model for $\zeta$ diverges from the simulations for frequencies smaller than 60 mHz. Considering three un-equal but constant arms for our semi-analytical response calculation for $\zeta$ extends the validity of the model to almost the entire \gls{lisa} required frequency range, while we still see a divergence between simulations and the model at very low frequencies below \SI{0.3}{\milli\hertz}. 
A preliminary study indicates that the mismatch is probably linked to the fact that we neglect the Sagnac effect in our model, i.e., that we assume the delays accross the two directions of each arm to be equal.

However, as the mismatch mostly occurs outside the required \gls{lisa} frequency band and the response of $\zeta$ remains sufficiently small compared to that of X inside the \gls{lisa} band down to \SI{0.1}{\milli\hertz} this does not significantly impact our conclusions.

\subsection{Combining Sagnac channels} \label{ssec:alpha-discussion}
\begin{figure}
	\centering
	\includegraphics[width=\columnwidth]{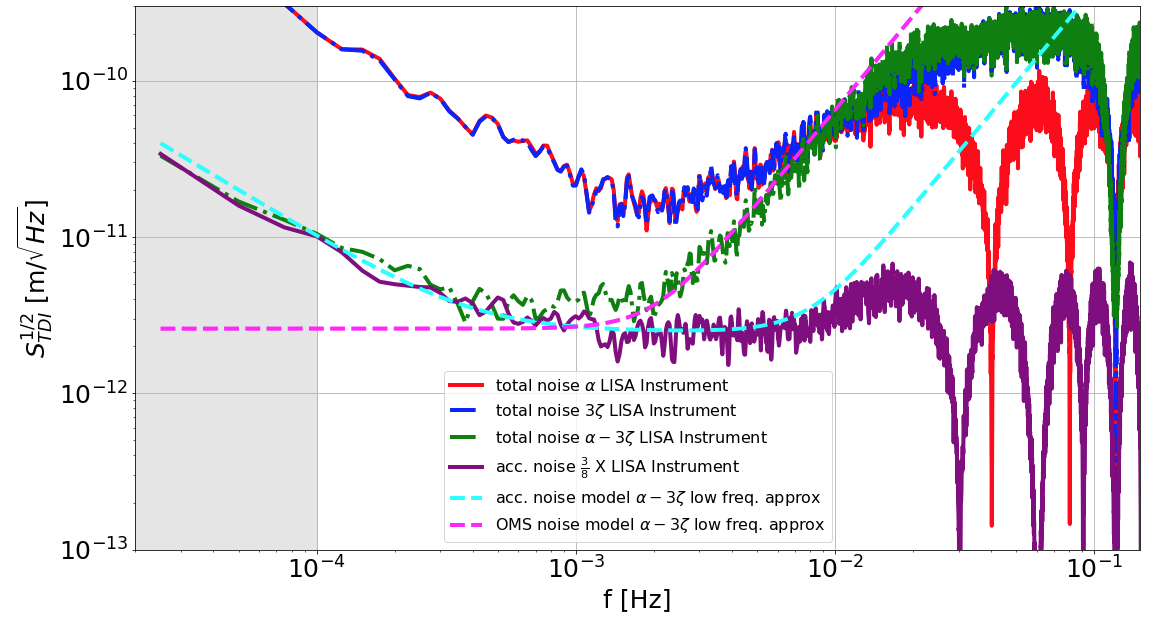}
	\caption{Amplitude spectral density of the TDI $\alpha$, 3$\zeta$ and $\alpha - 3\zeta$ for the acceleration noise and metrology noise. For TDI X only the amplitude spectral density of the acceleration noise is shown. The simulations use realistic ESA orbits included with \cite{lisa-orbits} while the models are derived assuming equal arms and considering only the low frequency component of the acceleration and metrology noise for the TDI combination $\alpha - 3\zeta$.\label{alphazeta}}
\end{figure}

First, let us consider the apparent possibility to use $\zeta$ to characterize and subtract the noise in $\alpha$. As discussed in~\cref{sec:disp}, the \gls{oms} noise contributions in $\alpha$ and $\zeta$ fulfill $ \alpha_{oms}( \tau)\approx 3 \zeta_{oms}( \tau)$ at low frequencies. This implies that subtracting  $3\zeta$ from $\alpha$ allows you to remove the common  \gls{oms} noise.
In~\cref{alphazeta} we report the simulations of the  \gls{tm} acceleration noise plus  \gls{oms} noise for $\alpha-3\zeta$ compared to the respective analytical models, which confirms what was predicted by the analytical calculation. What is then left as dominant noise source at low frequencies is a combinations of the following four  \glspl{tm}:
\begin{equation}
	\begin{split}
	[\alpha_{oms,g} & - 3 \zeta_{oms,g}] (t) \\
	& \approx 6 T^2 \Big({x^g_{12}}''(t)-{x^g_{13}}''(t) +{x^g_{21}}''(t)-{x^g_{31}}''(t)\Big).\label{alpz}
	\end{split}
\end{equation}
We notice that~\cref{alpz} is equal, up to a constant factor, to the low frequency \gls{tm} acceleration noise of the \gls{tdi} X channel (see \cref{eq:X_apprx}). This means that using the null channel $\zeta$ to reduce the noise, with the purpose of retrieving the  \gls{gw} signal in $\alpha$, gives you back the channel X, which is sensitive to \glspl{gw} as well as to the  \gls{tm} acceleration noise\footnote{As a remark, instead of utilizing $\zeta$ to remove the excess OMS noise in $\alpha$, one could also construct the optimal channels A and E out of the Sagnac variables, in which the dominant OMS noise terms also cancel. This follows readily from the result stated in~\cref{foot:3alphazeta} that the Sagnac channels fulfill $\alpha \approx \beta \approx \gamma$ for low-frequency OMS noise. Thus, OMS noise is cancelled to first order in both $A \simeq \alpha - \gamma$ and $E \simeq \alpha - 2\beta + \gamma$, giving these channels the same sensitivity as their Michelson-equivalents.
}. We therefore focus on just the X and $\zeta$ channels in the following.

\subsection{Upper limit on instrumental noise in X}
Let us now consider the expression for the \gls{oms} noise and \gls{tm} acceleration noise for \gls{tdi} $\zeta$ and X without considering the presence of \gls{gw} signals in our data. To put an upper limit on the instrumental noise in X we are looking for a frequency dependent factor $F$ such that $F S_{\zeta}\geq S_{X}$, which implies
\begin{equation}
F (S_{\zeta_{oms}} + S_{\zeta_{g}}) \geq S_{X_{oms}} + S_{X_g}.\label{eq:upb}
\end{equation}
Referring to~\cref{eq:S_dispX,eq:S_dispzeta,eq:S_gX,eq:S_gzeta}, this means
\begin{widetext}
\begin{equation}
 F T_{\zeta_{oms}} \sum_{ij \in \mathcal{I}_\zeta} S_{\text{oms}_{ij}} + F T_{\zeta_{g}} \sum_{ij \in \mathcal{I}_\zeta} S_{g_{ij}}^{disp}  \geq T_{X_{oms}} \sum_{ij \in \mathcal{I}_X} S_{\text{oms}_{ij}}  + T_{X_{g}} \left(\left(S_{g_{12}}^{disp}+S_{g_{13}}^{disp}\right) \cos ^2(\tau  \omega )+S_{g_{21}}^{disp}+S_{g_{31}}^{disp}\right). \label{eq:noise-upper-limit-start}
\end{equation}
Since  $\mathcal{I}_X \subset \mathcal{I}_\zeta$ and $S_{\text{oms}_{ij}}$ and $S_{\text{g}_{ij}}$ are strictly positive, we have
\begin{equation}
	\sum_{ij \in \mathcal{I}_\zeta} S_{\text{oms}_{ij}} \geq \sum_{ij \in \mathcal{I}_X} S_{\text{oms}_{ij}} 
\end{equation}
and further considering that  $\cos ^2(\tau \omega) \leq 1$ we get
\begin{equation}
	 \sum_{ij \in \mathcal{I}_\zeta} S_{g_{ij}}^{disp} \geq \left(\left(S_{g_{12}}^{disp}+S_{g_{13}}^{disp}\right) \cos ^2(\tau  \omega
	 )+S_{g_{21}}^{disp}+S_{g_{31}}^{disp}\right).
\end{equation}
We see that $F (S_{\zeta_{oms}} + S_{\zeta_{g}}) \geq S_{X_{oms}} + S_{X_g}$ is valid as long as
\begin{equation}
	F \geq T_{X_{oms}} /T_{\zeta_{oms}} \qquad \mathrm{and} \qquad 	F\geq T_{X_{g}}/ T_{\zeta_{g}} .
\end{equation}
We can therefore define our noise estimate factor as 
\begin{equation}
F = \mathrm{Max}( T_{X_{oms}} /T_{\zeta_{oms}} ,T_{X_{g}}/ T_{\zeta_{g}} ) = 256 \cos ^4\left(\frac{\omega \tau}{2}\right) \cos ^2(\omega \tau).
\end{equation}
\end{widetext}
Note that by inspection of~\cref{eq:S_gX,eq:S_gzeta,eq:S_dispX,eq:S_dispzeta}), we find the ratio $T_{X_{g}}/ T_{\zeta_{g}}$ to be dominant at all frequencies, which allows us evaluate the maximum in the previous equation.
\begin{figure*}
	\includegraphics[width=\columnwidth]{./figure/upperlimit.png}
	\includegraphics[width=\columnwidth]{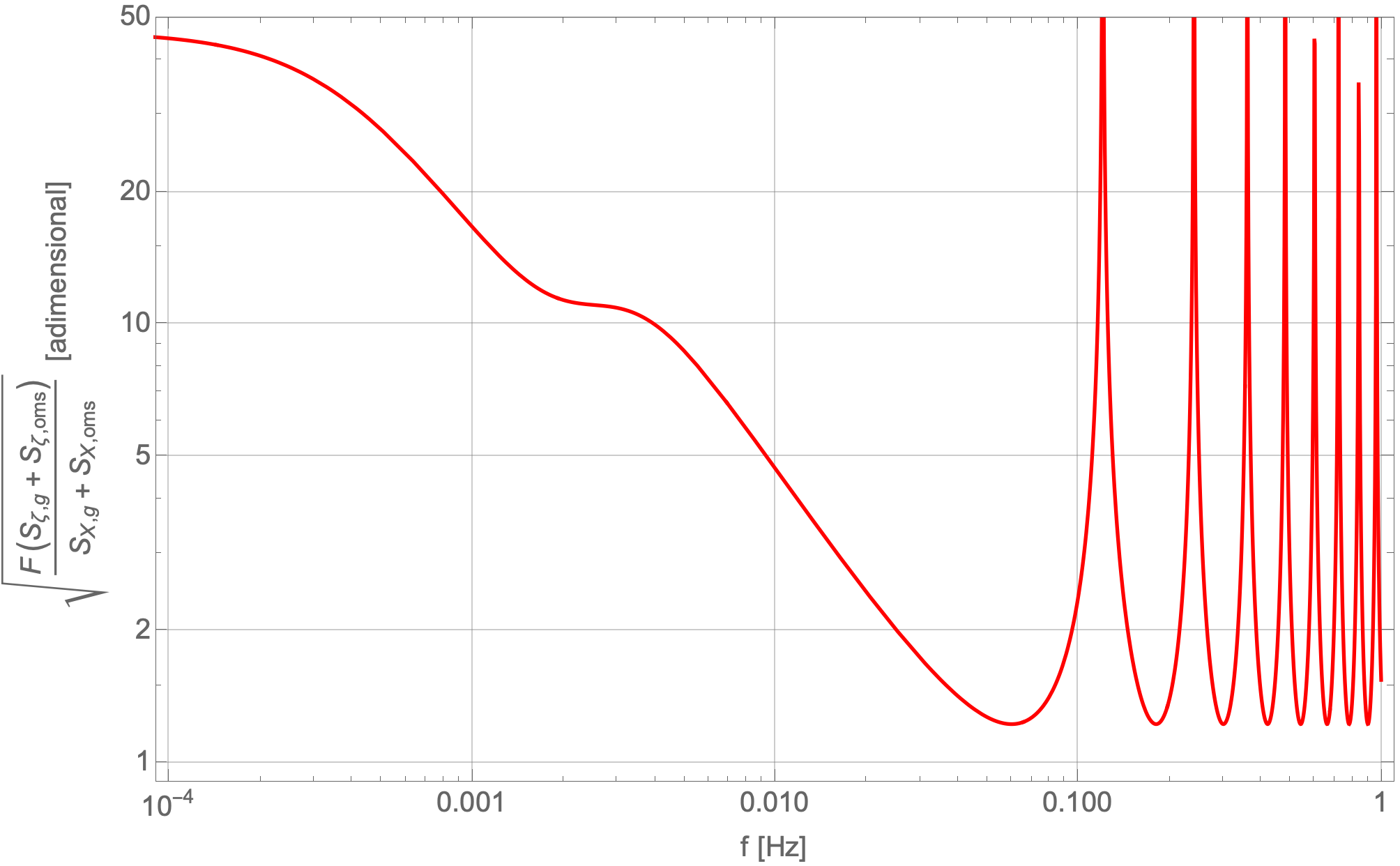}
	\caption{Left: Upper limit on the instrumental noise in X derived from the noise observed in $\zeta$, for the case where all TM acceleration and optical metrology noises are at the levels specified by~\cref{eq:tmacc,eq:readout}, respectively. Moreover, the dotted yellow and blue curves show the estimate one could put on just the OMS or TM noise, respectively. Right: Ratio between actual noise in X and the overall upper limit, in amplitude assuming required levels of TM acceleration and OMS noise. \label{fig:noise-upper-limit}}
\end{figure*}

We show in~\cref{fig:noise-upper-limit}, in the left plot, the overall noise upper limit $F S^{noise}_\zeta$ obtained in this way next to the actual noise $S^{noise}_X$ in X. In addition, we show the two individual upper limits we would obtain for just the  \gls{oms}  noise and just the \gls{tm} acceleration noise by considering only the contribution of $\frac{T_{X_{oms}} }{T_{\zeta_{oms}}} S^{noise}_\zeta$  and $\frac{T_{X_{g}}}{ T_{\zeta_{g}} }S^{noise}_\zeta $, respectively. 

Inspecting the right plot of~\cref{fig:noise-upper-limit}, we observe that (assuming noise at the required levels) the upper limit on the instrumental noise in X posed by $\zeta$ is up to a factor 50 in amplitude above the actual noise level, in particular at low frequencies. This results in a rather weak upper limit, reflecting \gls{oms} noise in a frequency band where only \gls{tm} acceleration noise is relevant. At high frequencies, on the other hand, where both X and $\zeta$ are dominated by \gls{oms} noise, the estimate is significantly more stringent, and stays below a factor 2 in amplitude from 25 to \SI{100}{\milli\hertz}.

We want to underline that the derivation of the expression $F S^{noise}_\zeta$ for the noise upper limit does not rely on any assumptions on the actual noise levels of the individual \gls{tm} and \gls{oms} noise terms, as only sums over all \gls{tm} and \gls{oms} channels affects~\cref{eq:noise-upper-limit-start}. Additionally, it could be evaluated at any time, and is therefore robust against non-stationarity of the noise. However, this upper limit does rely on our assumptions on noise correlations made in~\cref{sec:instrument-model}, and the particular outcome we show in~\cref{fig:noise-upper-limit} reflects the nominal values assumed for the level of \gls{tm} acceleration and \gls{oms} noise.

\subsection{Upper and lower limits on a SGWB}
We now additionally consider the presence of possible \glspl{sgwb} in our data, on which we can put lower and upper bounds as follows. As before for the instrumental noise, we will remain agnostic to the spectral shape and amplitude  of the \gls{sgwb}. We do however assume to know the response function of the different \gls{tdi} channels, which we compute as described in \cref{sensitivitycom} for the case of an isotropic \gls{sgwb}.

In the presence of such a \gls{sgwb} we can introduce
\begin{equation}\label{eq:total-Sxh}
	S^{meas}_X = S^{noise}_X +S_{X_h},
\end{equation}
as the combination of instrumental noise and GW signal that we can actually measure in the TDI X channel. 

We remind that~\cref{eq:def-Sxh} together with~\cref{eq:total-Sxh} immediately allows us to put an upper bound on a possible \glspl{sgwb},
\begin{equation} \label{eq:upperbound-X}
	S_h \leq S^{meas}_X / T_{X_h},
\end{equation}
as any model predicting a higher value of $S_h$ would be incompatible with our measurements $S^{meas}_X$.

To put a lower bound on $S_h$ based on our data (i.e., claim a detection), we can make use of our previously derived upper bound $F S^{noise}_\zeta$ on the instrumental noise in X.

To this end, as $\zeta$ is not perfectly insensitive to \glspl{gw} (cf.~\cref{fig:sens}), we need to define
\begin{equation}
	S^{meas}_\zeta = S^{noise}_\zeta +S_{\zeta_h}.
\end{equation}

The upper bound (cf.~\cref{eq:upb}) on the instrumental noise now becomes $F (S^{meas}_\zeta - S_{\zeta_h} )\geq S^{noise}_X$, which allows us to write
\begin{equation}
	S^{meas}_X \leq F (S_\zeta^{meas} - S_{\zeta_h} ) +S_{X_h}  ,
\end{equation}
where we simply add $S_{X_h}$ on both sides of the previous inequality and consider the definition of $S^{meas}_X$.
Then, considering~\cref{eq:def-Sxh} this implies the lower bound
\begin{equation}
	S_{h} \geq \frac{S^{meas}_X - F S^{meas}_{\zeta}}{T_{X_h} - F T_{\zeta_h}}, \quad \text{valid if $T_{X_h} > F T_{\zeta_h}$}.\label{eq:lowerbound}
\end{equation}%
Note that the right-hand side of this equation can be negative even if there is a \gls{sgwb}, in which case it is compatible with $S_h = 0$ and we cannot claim a detection of a \gls{sgwb}. On the other hand, if it is positive, this would indicate presence of a \gls{gw} background, at least in the assumptions used here.\\

\begin{figure*}
	\includegraphics[width=0.49\textwidth]{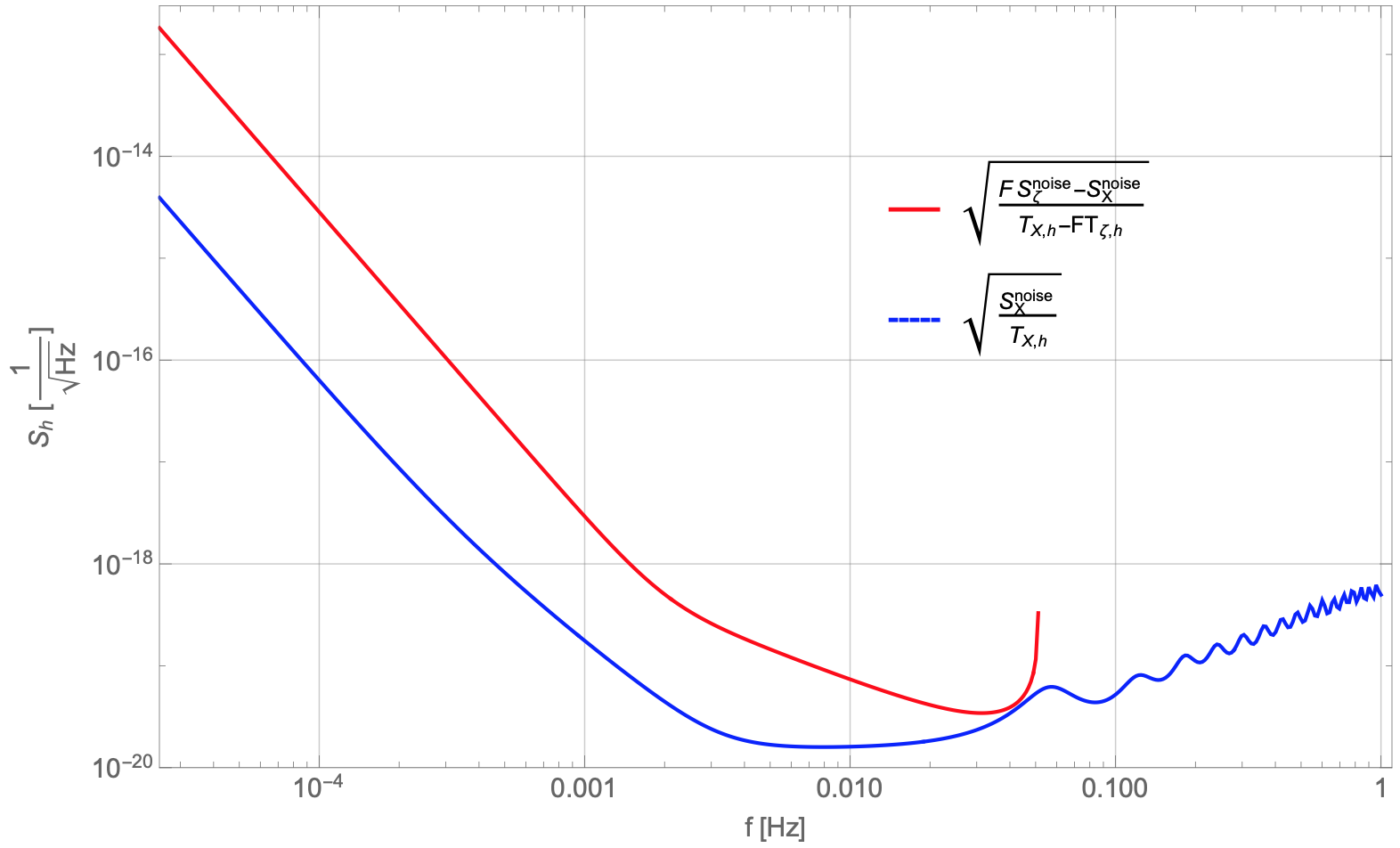}
        \includegraphics[width=0.47\textwidth]{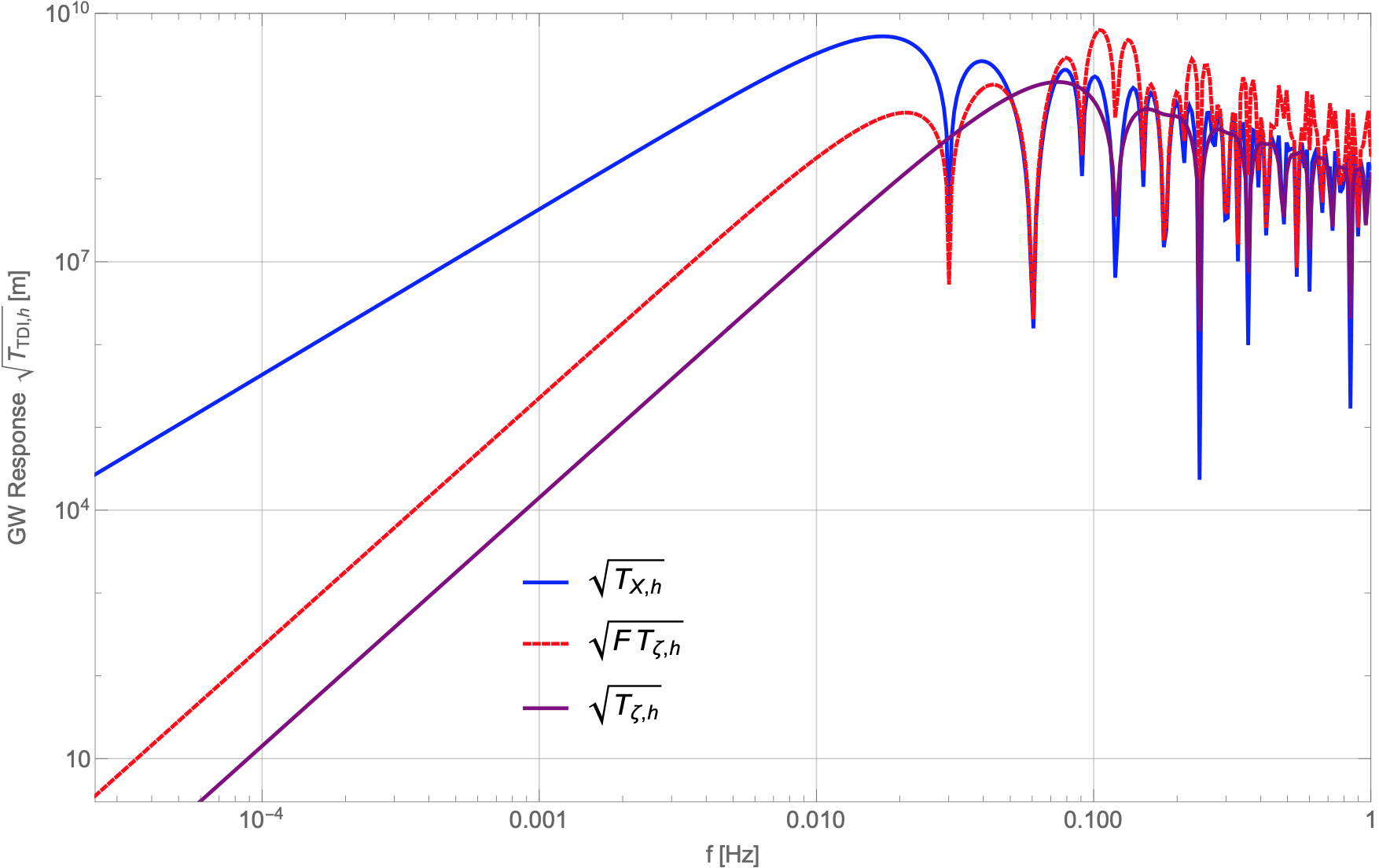}
	\caption{Left: Comparison of the detection threshold derived from $\zeta$ (red curve) and the upper limit given by X (blue curve), both expressed as relative armlength change (strain) assuming OMS and TM acceleration noise at the requirements level. Right: GW response functions of TDI X, $F S_{\zeta}$ and $S_{\zeta}$, computed assuming 51 stochastic GW sources isotropicaly distributed over the sky. \label{fig:GW-upper-limit}}
\end{figure*}

Assuming that we only consider the frequency range in which~\cref{eq:lowerbound} is valid, i.e., if $T_{X_h} > F T_{\zeta_h}$, the right-hand side of~\cref{eq:lowerbound} will be positive if $S^{meas}_X > F S^{meas}_{\zeta}$, which means we are able to identify a \gls{sgwb} if
\begin{subequations}
\begin{align}
	S^{noise}_{X} +  T_{X_h}S_h &> F(	S^{noise}_{\zeta} + T_{\zeta_h}S_h) \\
	\iff	 S_h &> \frac{F S^{noise}_{\zeta} - S^{noise}_{X}}{T_{X_h} - F T_{\zeta_h}}. \label{eq:detection-threshold}
\end{align}
\end{subequations}
\Cref{eq:detection-threshold} allows us to define a detection threshold assuming known noise levels, as depicted in the left plot of~\cref{fig:GW-upper-limit}, where it is shown alongside the upper limit defined in~\cref{eq:upperbound-X} in case we don't have a \gls{gw} background. Note that both these quantities now apply to the fundamentally unknowable  instrumental noise levels, and cannot be evaluated from the raw data.
We remind that the scaling factor $F$ used in this derivation was computed in the equal-arm assumption. While we showed in~\cref{ssec:analytical-model-and-simulations} that the equal-arm noise models are generally valid across most of the \gls{lisa} frequency band, we expect them to diverge in small frequency bands around the zeros of the \gls{tdi} transfer function\footnote{For example, the first zero of the second generation Michelson variable lies at roughly \SI{30}{\milli\hertz}. Assuming the arms of the constellation to be mismatched by 1 percent, the equal arm model is accurate to within 90 percent in a bandwidth of roughly \SI{1}{\milli\hertz} around this zero.}. 

This issue might be circumvented by using a different set of second generation \gls{tdi} variables which lack zeros at such low frequencies, as for example described in \cite{hartwig2022characterization}.
 \section{Conclusion\label{sec:conclusion}}
 
The \gls{lisa} data analysis, particularly in the search for a \gls{sgwb}, should be as robust as possible to ignorance of the noise model and to variations of the noise from the different components of the instrumental setup. 

It will likely be impossible to accurately predict and faithfully model the instrumental noise performance pre-flight, such that efforts to characterize the noise based on in-flight observables should be exploited as much as possible.
We present here how one can use the $\zeta$ channel to estimate the level at which two of the main noise sources, the uncorrelated \gls{tm} acceleration and \gls{oms} noise, will affect the \gls{gw} sensitive X channel. This is a rather conservative estimate, in the sense that it assumes nothing about actual instrumental noise levels, homogeneity between different \gls{tm} acceleration or \gls{oms} noise terms and noise stationarity. However, there are potential limits due to our assumptions on noise correlations, as $\zeta$ is highly insensitive to correlated noise entering both single-link measurements on-board a single spacecraft, while X is not.

We show that using $\zeta$ we estimate the noises under consideration in X within a factor 2 in amplitude in the band from \SI{25}{\milli\hertz} to \SI{100}{\milli\hertz}, while this estimate worsens to within a factor \num{50} in amplitude at the lowest frequencies (assuming the instrumental noise levels from the requirement). We can use this upper bound on the instrumental noise to compute a lower bound on the \gls{gw} background needed to explain the overall observed \glspl{psd} of both $\zeta$ and X. To this end, both the response of X to gravitational waves, $T_{X,h}$, and that of our instrumental noise estimate to gravitational waves, $F T_{\zeta,h}$, have to be considered. While $F T_{\zeta,h}$ is strongly subdominant to $T_{X,h}$ at low frequencies, this relationship is inverted at high frequencies, such that the lower bound becomes less stringent than one would expect from the performance of the noise estimate alone, and eventually becomes invalid. As visible in the right plot of~\cref{fig:GW-upper-limit}, we have $T_{X_h} >F  T_{\zeta_h}$ only up to around \SI{50}{\milli\hertz}. 

Note that the fact that the noise estimate $\zeta$ provides at low frequencies is a factor 50 above the actual instrumental noise in X implies that, within the assumption of this study, we could identify a \gls{sgwb} only if it were significantly larger than the \gls{tm} acceleration noise expected to dominate X at these frequencies\footnote{Such a strong SGWB could potentially bury the coherent signals associated with most of the LISA science case in gravitational noise.}.

Still, even assuming the nominal instrumental noise levels, this lower bound would allow to detect big stochastic backgrounds in a large part of the frequency band. 
Given the large uncertainties in the range of possible stochastic background levels \cite{caprini:hal-01703601}, including spectral shape and amplitude, as well as the demonstrated difficulty in predicting instrument noise, the results shown here might proof useful.  As such, our paper addresses the idea of simultaneous signal plus noise measurement, and shows the limit of achieving this with the TDI null channel.

Note that while our approach is agnostic to the noise levels, the predicted performance is computed assuming \gls{oms} noise to be exactly at the required noise levels (which includes a strong low-frequency relaxation), but, especially at low frequency, this noise has high uncertainty. If the actual hardware turns out to perform better in-flight than what can be demonstrated on-ground, the estimate would consequently improve. For example, earlier studies which performed similar estimates (e.g., \cite{Tinto:2001ii}, \cite{Hogan_2001}) assumed the \gls{oms} noise to be white across the whole frequency band, and came to the conclusion that we can make a better use of the null channels at low frequencies to estimate the \gls{sgwb}. We remark that for the \gls{oms} noise in $\zeta$ to reach the same level as the \gls{tm} acceleration noise (limiting the X channels at low frequencies) would require order of magnitude improvements in the performance of the \gls{oms}.

We want to reinforce that the upper and lower bounds we compute here are agnostic to the actual instrument performance and don't rely on any model of the individual noise spectral shapes or stationarities. This is in contrast to some other results in the literature (see for instance \cite{Adams_2010,Adams_2014, Caprini_2019, Flauger_2021,https://doi.org/10.48550/arxiv.2201.10902}), which showed it is possible to put significantly more stringent bounds on the noise assuming stationarity over the whole mission duration and a fixed (and known) noise shape which only depends on a single amplitude parameter. If indeed such a priori knowledge of the noise level and shape were possible, it would be possible to resolve \gls{sgwb} even below the threshold of the instrument noise.

The results presented here demonstrate the necessity of using realistic assumptions on the prior knowledge of the instrumental noise, noise correlations and stationarity.  It is important to consider that the data analysis pipelines in LISA operations will likely rely on some model 
for the noise (even if Bayesian techniques for parameter's estimation with unknown noise has been introduced in literature, 
see for instance \cite{PhysRevD.90.042003} ). Although procedures like those described in this manuscript do not translate naturally into a Bayesian data analysis framework we believe they might still proof useful to cross-check and interpret the results from a full Bayesian analysis, given the large number of parameters such a procedure has to determine. Additionally, the lower and upper bounds provided from our method could be used as priors in a Bayesian framework.

Further studies should be performed to quantify the real impact this has on achieving the \gls{lisa} science objectives to detect \glspl{sgwb}.

To conclude, we reiterate that this study is limited in that we only considered the two main classes of noise, \gls{tm} acceleration and \gls{oms} noise, and that we further assume that these are fully uncorrelated for the six \glspl{tm} and six one-way optical metrology links. Follow-up studies could investigate other known noise sources with different correlation properties, such as sideband modulation noise \cite{Hartwig_2021} or \gls{ttl} couplings \cite{PhysRevD.106.042005}, to verify to which extend the results presented here hold for such noises. Furthermore, we only considered here the case of an isotropic \gls{sgwb} for simplicity. Any anisotropic \gls{sgwb}, such as the expected foreground from galactic binaries, will have an annual modulation in the response function, which might help to distinguish it better from the instrumental noise. We want to remark that some instrumental noises might also show annual  modulations due to the position of the LISA satellites along the orbit, which one should account for when studying this scenario.

\section{Acknowlegement}
We thank the LISA simulation working group, in particular Jean-Baptiste Bayle,  Quentin Baghi, Maude Le Jeune, Arianna Renzini and  Martin Staab for developing the tools used for the simulations. We also thank the anonymous referee for the useful comments on improving the manuscript. M.M and O.H. want to thank Antoine Petiteau, Mauro Pieroni, Marc Lilley and Jonathan Gair for interesting discussions regarding this topic. M.M, S.V, D.V. and W.J.W. thank the LISA Trento group for the fruitful discussion and the Istituto Nazionale di Fisica Nucleare (INFN) for supporting this work. This work was funded by the Agenzia Spaziale Italiana (ASI), Project No. 2017-29-H.1-2020 "Attività per la fase A della missione LISA".
O.H. gratefully acknowledges support by the Centre national d'études spatiales. This work was supported by the Programme National GRAM of CNRS/INSU with INP and IN2P3 co-funded by CNES.
M.M. gratefully acknowledge support by the Deutsches Zentrum fur Luft- und Raumfahrt (DLR) with funding from the Bundesministerium fur Wirtschaft und Technologie (Project Ref. Number 50 OQ 1801)
\appendix
\section{Time shift operators}
\label{sec:notations}
We define the following notations related to time-shift operators and \gls{tdi} combinations \cite{hartwig2022characterization}:
\paragraph*{Delay operator: }
 \begin{equation}
D_{ij}\eta(\tau) = \eta(\tau - d_{ij}(\tau)).
 \end{equation}
	Given a time of reception $\tau$ of a beam on spacecraft $i$, evaluates the measurements $\eta$ (we dropped the double index for simplicity) of that beam at the time of emission at spacecraft $j$, which we write as $\tau-d_{ij}(\tau)$. Note that depending on what frame $\eta(\tau)$ is defined in, the computation of $d_{ij}$ can include a change in reference frames and clock offsets, as discussed in \cite{case1-arxiv}.
\paragraph*{Advancement operator: } 
\begin{equation}
A_{ij} \eta(\tau) = \eta(\tau + a_{ij}(\tau)).
\end{equation}	
	Given a time of emission $\tau$ of a beam from spacecraft $j$, evaluates the phase $\eta$ of that beam at the time of reception on spacecraft $i$, which we write as $\tau+a_{ij}(\tau)$. This is the inverse operation to that of the delay operator, such that we have the identity $A_{ij}D_{ji} \eta(t) = D_{ij}A_{ji} \eta(t) =  \eta(t)$.
\paragraph*{Multiple Delay operators: }
\begin{equation}
D_{ij} D_{jk}\eta(\tau) = \eta(\tau-d_{ij}(\tau) - d_{jk}(\tau-d_{ij}(\tau)).
\end{equation} 
\paragraph*{Multiple Delay and Advancement operators: } 
\begin{equation}
	\begin{split}
		A_{n i}  D_{ij} &D_{j k}\eta(\tau) = \eta\Big(\tau+ a_{ni}(\tau)-d_{ij}(\tau+ a_{ni}(\tau)) \\
		& \quad -d_{jk}(\tau+ a_{ni}(\tau)-d_{ij}(\tau+ a_{ni}(\tau)))\Big).
	\end{split}
\end{equation}
 Only the delays $d_{ij}(\tau)$ are directly accessible from the \gls{lisa} measurements. The advancements $a_{ij}(\tau)$ can be computed from them by iteratively solving
 \begin{equation}
 	a_{ij}(\tau) = d_{ji}(\tau + a_{ij}(\tau)),
 \end{equation}
 which directly follows from $A_{ij}D_{ji} \eta(t) =  \eta(t)$.
\section{TM acceleration and displacement noise models\label{ssec:tm-noise}}

Following the convention that $\vec{L}_{ji}$ is the link vector from the emitting satellite OBj to the receiving one OBi, and $\vec{g}_{i}$ the OBi acceleration relative to its inertial reference frame, we can define the acceleration of OBi that points towards OBj, at time t, as:
\begin{equation}
g_{ij}(t) \equiv \vec{g}_{i}(t)\cdot \hat{L}_{ji}.
\end{equation}
Then, the relative acceleration $\Delta g_{\textrm{single-link}}(t)$, between the two free-falling \glspl{tm} along the line of sight of the unit vector $\hat{L}_{ji}$, at time $t$ on $OB_i$, can be computed as:
\begin{align}
\Delta{g_{\textrm{single-link}}}(t)&= (\vec{g}_{i}(t) -\vec{g}_{j}(t-\tau))\cdot \hat{L}_{ji} \\
&  \equiv (g_{ij}(t)+g_{ji}(t-\tau)),
\end{align}
where we have use the approximation that: 
\begin{equation}
\vec{g}_{i}(t) \cdot \hat{L}_{ji} \approx -\vec{g}_{i}(t)\cdot \hat{L}_{ij}.
\end{equation}
We can estimate the \gls{psd} of $\Delta{g_{\textrm{single-link}}}$ under the assumption of uncorrelated but statistically equivalent acceleration noises for the two \glspl{tm} as: 
\begin{equation}
S_{\Delta{g_{\textrm{single-link}}}}(\omega) =  2S_{g_{ij}} ,
\end{equation}
where $ S_{g_{ij}}$ is the \gls{psd} of the single \gls{tm} acceleration noise.
To give an estimate of the \gls{oms} noise for the inter-spacecraft interferometer in a \gls{lisa} link, we should consider that it enters just at the time $t$ when we perform the measurement, as:
\begin{equation}
\Delta x_{\textrm{single-link}}(t) = x_{ij}(t).
\end{equation}
Here $ x_{ij}$ is the readout noise expressed in term of displacement at OBj that faces the far OBi. 

\section{\label{a:acc}Analytical computation in time domain of the acceleration noise and displacement noise for the TDI X, \texorpdfstring{$\alpha$}{alpha}, \texorpdfstring{$\zeta$}{zeta}}
We can compute how the \gls{tm} acceleration noise propagates through the \gls{tdi} X, $\alpha$ and $\zeta$, assuming equal and constant arm lengths as follows:
\begin{widetext}
\begin{subequations}
	\begin{align}
		\begin{split}	\label{Eq:X} 
		X_{g}( t)  = & g_{12}(t-8 \tau)-2 g_{12}(t-4 \tau)-g_{13}(t-8 \tau)+2 g_{13}(t-4 \tau)+2 g_{21}(t-7 \tau)  
		\\ 
		&  -2 g_{21}(t-5 \tau)-2 g_{21}(t-3 \tau)  +2 g_{21}(t-\tau)-2 g_{31}(t-7 \tau)+2 g_{31}(t-5 \tau) 
		\\ 
		&  +2 g_{31}(t-3 \tau)-2 g_{31}(t-\tau)+g_{12}(t)-g_{13}(t),
	    \end{split}
		\\
		\begin{split} \label{Eq:alpha} 
			\alpha_{g}( \tau) = &  g_{12}(t-6 \tau )-2 g_{12}(t-3 \tau )-g_{13}(t-6 \tau )+2 g_{13}(t-3 \tau )+g_{21}(t-5 \tau )-g_{21}(t-4 \tau ) 
			\\ 
			&  -g_{21}(t-2 \tau )+g_{21}(t-\tau )+g_{23}(t-5 \tau )-g_{23}(t-4 \tau )
			\\ 
			& -g_{23}(t-2 \tau )+g_{23}(t-\tau ) -g_{31}(t-5 \tau)+g_{31}(t-4 \tau )+g_{31}(t-2 \tau )-g_{31}(t-\tau) 
			\\ 
			& -g_{32}(t-5 \tau )+g_{32}(t-4 \tau )+g_{32}(-2 \tau
			)-g_{32}(t-\tau )+g_{12}(0)-g_{13}(0), 
		\end{split}
	\\
\begin{split}\label{Eq:zeta}
		\zeta_{g}( t) = &  g_{12}(t-2 \tau )-2 g_{12}(t-\tau )-g_{1,3}(t-2 \tau )+2 g_{13}(t-\tau )   -g_{21}(t-2 \tau )+2 g_{21}(t-\tau )
		\\ 
		&+g_{23}(t-2\tau )-2 g_{2,3}(t-\tau )+g_{3,1}(t-2 \tau ) -2 g_{31}(t-\tau )   
		\\ 
		& -g_{32}(t-2 \tau )+2 g_{32}(t-\tau)+g_{12}(t)-g_{13}(t)-g_{21}(t)+g_{23}(t)+g_{31}(t)-g_{32}(t).
\end{split}		
	\end{align}
\end{subequations}
Following the same assumption we used for computing the \gls{tm} acceleration noise, we can also compute how the \gls{oms} noise enters in the above mentioned \gls{tdi} channels:
\begin{subequations}
	\begin{align}
		\begin{split} \label{Eq:Xoms}
		X_{oms}( t)  =&  x_{12}(t-6 \tau )-x_{12}(t-4 \tau )-x_{12}(t-2 \tau )-x_{13}(t-6 \tau )+x_{13}(-4 \tau )  
		 \\ 
		 & +x_{13}(t-2 \tau) +x_{21}(t-7 \tau )-x_{21}(t-5 \tau ) -x_{21}(t-3 \tau )+x_{21}(t-\tau )  
		 \\ 
		 &  -x_{31}(t-7 \tau )+x_{31}(t-5 \tau)  +x_{31}(t-3 \tau )-x_{31}(t-\tau )+x_{12}(t)-x_{13}(t), 
	\end{split} 
\\
\begin{split}\label{Eq:alphaoms}
		\alpha_{oms}(t)  =&  -x_{12}(t-3 \tau )+x_{13}(t-3 \tau )+x_{21}(t-5 \tau )-x_{21}(t-2 \tau ) 
		\\ 
		& -x_{23}(t-4 \tau )+x_{23}(t-\tau)   -x_{31}(t-5 \tau )+x_{31}(t-2 \tau )+x_{32}(t-4 \tau )
		\\ 
		& -x_{32}(t-\tau )+x_{12}(t)-x_{13}(t), 
\end{split}		
\\
\begin{split}\label{Eq:zetaoms}
\zeta_{oms}(t) = & -x_{12}(t-\tau )+x_{13}(t-\tau )+x_{21}(t-\tau )-x_{23}(t-\tau )-x_{31}(t-\tau ) 
\\ 
& +x_{32}(t-\tau)+x_{12}(t)-x_{13}(t)-x_{21}(t)+x_{23}(t)+x_{31}(t)-x_{32}(t).
\end{split}
\end{align}
\end{subequations}
\end{widetext}
\section{Computation of the Sensitivity}\label{sensitivitycom}
Following \cite{Muratore:2021phd} and \cite{muratore2021time}, We consider stochastic sources with both plus and cross polarizations in their source frame. In the Solar System Barycenter (SSB), these will appear with $h_{+}(t,\textbf{r})$ and $h_{\times}(t,\textbf{r})$ given by
$h^{SSB}_{+}(t,\textbf{r}) = h_{+}(t,\textbf{r}) \cos(2 \psi)-h_{\times}(t,\textbf{r}) \sin(2 \psi)$ and $h^{SSB}_{\times}(t,\textbf{r}) = h_{+}(t,\textbf{r}) \sin(2 \psi) +h_{\times}(t,\textbf{r}) \cos(2 \psi)$, where $\psi$  is the polarization angle. 
 The sensitivity to GW sources coming from different directions is computed for each source considering the relative frequency shift that an incoming \gls{gw} causes on a \gls{lisa} link as for example given in \cite{BPH2021}. We then convert this frequency shift to an equivalent displacement. We computed both the case of three equal armlength and three unequal constant armlength. 
 Assuming that our signal is made of superposition of many \gls{gw} sources coming from different directions and with different polarizations, we can consider that the output of a TDI$_j$, given superpositions of $n$ plane waves is:
\begin{equation}
S_{{j}_h} = \sum_{i}^n T_{j_h}^i(\omega) S_{h_i}(\omega)\label{sh},
\end{equation}
where $S_{h_i}(\omega)$ is the \gls{psd} of the $i$'th \gls{gw} source expressed as dimensionless strain, and $T_{j_h}^i(\omega)$ is the absolute squared value transfer function for the $j$'th \gls{tdi}, including the conversion factor such that $S_{{j}_h}$ is in units of \si{\meter\squared\per\hertz}. 
Labelling the \gls{psd} of the \gls{tm} acceleration noise and \gls{oms} noise for each \gls{tdi} $j$ as $S_{{j}_g}$ and $S_{{j}_{oms}}$, respectively, the sensitivity of each \gls{tdi} combination is computed by renormalising the total instrument noise \gls{asd} by the \gls{gw} transfer function as:
\begin{equation}
S_{{j}_{n/h}} = \frac{S^{disp}_{{j}_g}+ S_{{j}_{oms}} }{RMS \{T_{j_h}^i(\omega)\}},
\end{equation}
where the $RMS\{\}$ denotes the root mean square over all sources $i$ and as before $S^{disp}_{{j}_g}$ is the \gls{tm} acceleration noise expressed as an equivalent displacement. \\
The response to a SGWB can also be written using a continous integral over the whole sky as for example shown in \cite{Smith_2019} and \cite{Flauger_2021}. The angular integral reported there is then evaluated numerically to get a result which is valid for the whole LISA frequency range. The computation reported in this paper is one possible method for numerically approximating the result of the continous integral by replacing it with a sum over discrete stochastic sources from different directions. Indeed, in the limit of an infinite number of sources, this converges to the same integral.

\bibliography{bibliography}

\end{document}